\documentclass[ALICE,manyauthors]{cernphprep}
\usepackage{hyperref}
\usepackage{lineno}
\usepackage[T1]{fontenc}
\usepackage[english]{babel}
\usepackage{graphicx}
\usepackage{verbatim}
\usepackage{amsfonts}
\usepackage{makeidx}
\usepackage{color}
\usepackage{amsmath}
\usepackage{lineno}
\usepackage{epsfig}
\usepackage{epstopdf}
\usepackage{hyperref}
\usepackage{cite}
\usepackage{textcomp}
\usepackage{multirow}
\usepackage{makecell}
\usepackage{booktabs}
\newcommand{\pp}{pp}

\newcommand{\sqrts}{\sqrt{s}}
\newcommand{\sqrtsNN}{\sqrt{s_{\rm \scriptscriptstyle NN}}}

\newcommand{\av}[1]{\left\langle #1 \right\rangle}

\newcommand{\MeV}{\mathrm{MeV}}
\newcommand{\GeV}{\mathrm{GeV}}
\newcommand{\TeV}{\mathrm{TeV}}

\newcommand{\gevc}{\mathrm{GeV}/c}

\newcommand{\fmc}{\mathrm{fm}/c}

\newcommand{\PbPb}{\mbox{Pb--Pb}~}
\newcommand{\pPb}{\mbox{p--Pb}~}
\newcommand{\AuAu}{\mbox{Au--Au}~}

\newcommand{\Ncoll}{N_{\rm coll}}
\newcommand{\Raa}{R_{\rm AA}}

\newcommand{\RpPb}{R_{\rm pPb}}
\newcommand{\RAA}{R_{\rm AA}}
\newcommand{\TAA}{T_{\rm AA}}
\newcommand{\pt}{p_{\rm T}}

\newcommand{\dNdy}{{\rm d}N/{\rm d}y}

\newcommand{\DtoKpi}{{\rm D}^0 \to {\rm K}^-\pi^+}
\newcommand{\DtoKpipi}{{\rm D}^+\to {\rm K}^-\pi^+\pi^+}
\newcommand{\DstartoDpi}{{\rm D}^{*+} \to {\rm D}^0 \pi^+}

\newcommand{\Dzero}{{\rm D^0}}
\newcommand{\Dzerobar}{{\overline{{\rm D}}\,^0}}
\newcommand{\Dstar}{{\rm D^{*+}}}

\newcommand{\Dplus}{{\rm D^+}}

\newcommand{\Ds}{{\rm D_s^+}}

\newcommand{\dEdx}{d$E/$d$x$}

\newcommand{\Jpsi}{{\rm J}/\psi}

\newcommand{\dNAAdpt}{{\rm d}N_{\rm AA}/{\rm d}\pt}

\newcommand{\dNdydpt}{{\rm d}^2N/{\rm d}y {\rm d}\pt}

\newcommand{\dsigmadpt}{{\rm d}\sigma_{\rm pp}/{\rm d}\pt}

\begin{document}%

\begin{titlepage}
\PHyear{2021}
\PHnumber{213}      
\PHdate{13 October}  
%

\title{Prompt $\Dzero$, $\Dplus$, and $\Dstar$ production in \\ Pb--Pb collisions at $\pmb{\sqrtsNN = 5.02~\TeV}$}
\ShortTitle{Prompt $\Dzero$, $\Dplus$, and $\Dstar$ $\RAA$ in Pb--Pb}   

\Collaboration{ALICE Collaboration\thanks{See Appendix~\ref{app:collab} for the list of collaboration members}}
\ShortAuthor{ALICE Collaboration} 

\begin{abstract}
The production of prompt $\Dzero$, $\Dplus$, and $\Dstar$ mesons was measured at midrapidity ($|y| <~$0.5) in \PbPb collisions at the centre-of-mass energy per nucleon--nucleon pair $\sqrtsNN = 5.02~\TeV$ with the ALICE detector at the LHC. The D mesons were reconstructed via their hadronic decay channels and their production yields were measured in central (0--10\%) and semicentral (30--50\%) collisions. The measurement was performed up to a transverse momentum ($\pt$) of 36 or 50~$\gevc$ depending on the D meson species and the centrality interval.
For the first time in \PbPb collisions at the LHC, the yield of $\Dzero$ mesons was measured down to $\pt =$ 0, which allowed a model-independent determination of the $\pt$-integrated yield per unit of rapidity ($\dNdy$). 
A maximum suppression by a factor 5 and 2.5 was observed with the nuclear modification factor ($\RAA$) of prompt D mesons at $\pt =$ 6--8$~\gevc$ for the 0--10\% and 30--50\% centrality classes, respectively.
The D-meson $\RAA$ is compared with that of charged pions, charged hadrons, and $\Jpsi$ mesons as well as with theoretical predictions. The analysis of the agreement between the measured $\RAA$, elliptic ($v_{2}$) and triangular ($v_{3}$) flow, and the model predictions allowed us 
to constrain the charm spatial diffusion coefficient $D_{s}$. Furthermore the comparison of $\RAA$ and $v_{2}$ with different implementations of the same models provides an important insight into the role of radiative energy loss as well as charm quark recombination in the hadronisation mechanisms.

\end{abstract}
\end{titlepage}
\setcounter{page}{2}

\section{Introduction}
\label{sec:introduction}
Ultra-relativistic heavy-ion collisions allow the study of quantum chromodynamics (QCD) at high energy density and temperature. 
According to lattice QCD (lQCD) calculations, these extreme conditions lead to a transition of hadronic matter into a strongly-interacting medium, called quark--gluon plasma (QGP), in which quarks and gluons are deconfined and chiral symmetry is partially restored~\cite{Collins:1975us, Karsch:2006xs, Borsanyi:2010bp, Borsanyi:2013bia, Bazavov:2011nk}. A QGP is formed and studied in high-energy heavy-ion collisions at the Relativistic Heavy Ion Collider (RHIC) and at the CERN Large Hadron Collider (LHC). The existing measurements indicate that the QGP behaves as a strongly-coupled low-viscosity liquid-like system~\cite{Busza:2018rrf} and its lifetime at the energy densities reached at the LHC is of the order of 10 $\fmc$~\cite{ALICE:2011dyt}.

Heavy quarks, like charm and beauty, are mostly produced in primary hard scattering processes between the partons of the incoming nuclei, which occur in the early stages of the collisions. 
The time scales for heavy-quark production ($\leq$ 0.07 $\fmc$ for ${\rm c}\overline{\rm c}$ and $\leq$ 0.02 $\fmc$ for ${\rm b}\overline{\rm b}$ pairs~\cite{Andronic:2015wma}) are shorter than the QGP formation time, which is about 0.3--1.5 $\fmc$ at LHC energies~\cite{Liu:2012ax}. Therefore, heavy quarks experience the full space--time evolution of the hot and dense QCD medium. 

During their propagation in the QGP, heavy quarks interact with the medium constituents, by exchanging energy and momentum via elastic~\cite{Thoma:1990fm, Braaten:1991jj, Braaten:1991we} and inelastic~\cite{Baier:1996sk, Gyulassy:1990ye} processes. Low-momentum heavy quarks (i.e.~$\lesssim$ 3--4~$\gevc$) mainly interact via elastic scatterings and are expected to acquire some collective behaviour of the system, like radial and anisotropic azimuthal flow, as a consequence of multiple interactions with the medium~\cite{Batsouli:2002qf,Greco:2003vf}. The typical momentum exchange in the interactions of heavy quarks with the medium is small compared to the charm-quark mass. Therefore, heavy quarks undergo Brownian motion in the medium characterised by many small-momentum kicks. The degree of heavy-quark thermalisation provides unique insight into the thermalisation process of the system. Substantial theoretical efforts were made to describe the charm-quark transport in a hydrodynamically expanding medium~\cite{Rapp:2018qla}. The relevant transport parameter is the spatial diffusion coefficient $D_s$, which does not depend on the precise value of the heavy-quark mass and, hence, is a medium property. It can be estimated via the comparison of the measured differential yields as a function of transverse momentum ($\pt$) and azimuthal distributions of D mesons with model predictions.

For high-$\pt$ quarks (i.e.~$\gtrsim$ 8--10~$\gevc$) the main manifestation of the interactions with the medium is energy loss, which can occur via collisional processes and gluon radiation~\cite{Prino:2016cni}, with the latter expected to be the dominant mechanism. 
In particular, the amount of radiative energy loss is influenced by the colour charge of the parton interacting with the medium. According to QCD calculations, gluons are expected to lose more energy than quarks, due to their stronger coupling to the medium~\cite{Gyulassy:1990ye,Baier:1996sk}. At LHC energies, light-flavour particles in the momentum interval 5 $<\pt<$ 20~$\gevc$ are expected to originate mostly from the fragmentation of gluons produced in hard scattering processes, while at higher $\pt$ the contribution of light-quark jets becomes relevant~\cite{Djordjevic:2013pba}. On the other hand, charm mesons provide an experimental tag for a quark parent at all momenta. Thus, the comparison of heavy-flavour hadron production with that of light-flavour particles at high $\pt$ can provide insight into this aspect.
In addition, the energy loss of quarks can be affected by several mass-dependent effects like the dead-cone effect, which reduces the small-angle gluon radiation for quarks with moderate energy-over-mass values~\cite{Dokshitzer:2001zm,Armesto:2003jh,Djordjevic:2003zk,Zhang:2003wk}. Similarly, collisional energy loss is predicted to depend on the quark mass and to be smaller for heavy quarks~\cite{vanHees:2005wb}.

Heavy-flavour hadron yields and momentum distributions can be influenced by the modification of the parton distribution function (PDF) due to initial state effects, like nuclear shadowing~\cite{Prino:2016cni}. It was also suggested that low-momentum heavy quarks could hadronise via recombination with other quarks from the medium, in addition to the fragmentation in the vacuum~\cite{Greco:2003vf, Andronic:2003zv}.

The aforementioned effects can be investigated with the measurement of the nuclear modification factor $\RAA$ of heavy-flavour hadrons. It is defined as the ratio of the $\pt$-differential production yield in nucleus--nucleus collisions ($\dNAAdpt$) and the production cross section in proton--proton collisions ($\dsigmadpt$) scaled by the average nuclear overlap function $\langle\TAA\rangle$
\begin{equation}
\label{eq:Raa}
R_{\rm AA}(\pt)=
{1\over \av{T_{\rm AA}}} {{\rm d} N_{\rm AA}/{\rm d}\pt \over
{\rm d}\sigma_{\rm pp}/{\rm d}\pt},
\end{equation}

where $\langle\TAA\rangle$ is defined as the average number of binary nucleon--nucleon collisions ($\langle\Ncoll\rangle$), which can be estimated via Glauber model calculations~\cite{Glauber:1970jm,Miller:2007ri,Loizides:2014vua}, divided by the inelastic nucleon--nucleon cross section.

Measurements of prompt D-meson production were performed by the ALICE Collaboration in \PbPb collisions at $\sqrtsNN =$ 2.76 TeV~\cite{Adam:2015sza} and 5.02~$\TeV$~\cite{Acharya:2018hre} in the 0--10\% and 30--50\% centrality classes for both collision energies and in the 60--80\% for the highest one. The D-meson yields show a suppression in central (0--10$\%$) and semi-central (30--50$\%$) collisions that reaches up to a factor of 5 and 2.5, respectively, at $\pt$ of 6--10 GeV/\textit{c}. It then decreases with increasing $\pt$, while the suppression factor is of 1.25 without a pronounced $\pt$ dependence in peripheral (60--80$\%$) collisions. The prompt $\Dzero$ production was also measured by the CMS Collaboration in the most central \PbPb collisions at $\sqrtsNN = 5.02~\TeV$~\cite{Sirunyan:2017xss} in the $\pt$ range 2--100~$\gevc$ and the result is in agreement with ALICE measurements.
Furthermore, also the STAR Collaboration measured the production of the prompt $\Dzero$ mesons in \AuAu collisions at $\sqrtsNN =$ 200 GeV~\cite{Adam:2018inb} and the comparison with this result can provide insight into the $\sqrts$ dependence of energy-loss effects.
Conversely, the D-meson nuclear modification factor measured by the ALICE Collaboration in \pPb collisions at $\sqrtsNN = 5.02~\TeV$~\cite{Acharya:2019mno}, where an extended QGP is not expected to form, is compatible with unity for the whole measured $\pt$ range.

Complementary information on the interactions of heavy quarks with the medium constituents and their possible thermalisation can be obtained from the measurements of the anisotropies in the D-meson azimuthal distributions, which are characterised in terms of the Fourier coefficients~\cite{Voloshin:1994mz,Poskanzer:1998yz}. In particular, the second order coefficient $v_{2}$, called elliptic flow, is defined as  $v_2=\langle \cos (\mathrm{2}\varphi)\rangle$ where $\varphi$ is the azimuthal angle.
Recently, the ALICE Collaboration reported the latest measurements of prompt D-meson azimuthal anisotropic flow~\cite{Acharya:2020pnh} in \PbPb collisions at $\sqrtsNN = 5.02~\TeV$.

In this paper, the measurements of the $\pt$-differential yields and nuclear modification factors of prompt $\Dzero$, $\Dplus$, and $\Dstar$ mesons (i.e.~produced via the hadronisation of charm quarks or from the decays of excited charmed hadron states) and their charge conjugates, in central (0--10$\%$) and semicentral (30--50$\%$) \PbPb collisions at $\sqrtsNN = 5.02~\TeV$ are reported.
The statistical and systematic uncertainties have been reduced due to the \PbPb data sample collected with the ALICE detector at the end of 2018, that is larger by a factor eight (four) for central (semicentral) collisions with respect to the data sample collected in 2015 used for the previous publication~\cite{Acharya:2018hre}, and to the production cross section measured in pp collisions at the same centre-of-mass energy~\cite{Acharya:2019mgn}.
The larger data sample allowed the reduction of the statistical uncertainties and the extension of the $\pt$ reach of the measurements down to $\pt =$ 0 for $\Dzero$ mesons. Moreover, the previous measurements~\cite{Acharya:2018hre} were affected by the uncertainty on the $\sqrts$ scaling of the pp production cross section. The new measurements do not include such uncertainty since they use the measured pp production cross section at the same collision energy~\cite{Acharya:2019mgn}.

The paper is structured as follows. The data sample and the experimental apparatus are briefly introduced in Section~\ref{sec:detector}. The D-meson reconstruction procedure and the corrections applied to the raw yields are presented in Section~\ref{sec:analysis}, while the estimation of the systematic uncertainties is described in Section~\ref{sec:systematics}. The results are presented in Section~\ref{sec:results}, together with a comparison with the charged-pion, proton, and prompt and non-prompt $\Jpsi~\Raa$, along with model calculations.
In addition, a simultaneous comparison of the measured $\RAA$ and elliptic flow $v_2$~\cite{Acharya:2020pnh} at low and intermediate $\pt$ with transport-model calculations is discussed, together with an estimate of the spatial diffusion coefficient. This comparison will provide insights into the participation of the charm quark in the collective expansion of the medium~\cite{Moore:2004tg}, as well as on the role of the different hadronisation mechanisms. Conclusions and perspectives are drawn in Section~\ref{sec:conclusions}.

\section{Detector and data sample}
\label{sec:detector}

The ALICE experimental apparatus includes several detectors for particle tracking and identification as well as electromagnetic calorimeters at midrapidity ($|\eta|<0.9$), a forward muon spectrometer ($-4<\eta<-2.5$), and a set of forward and backward detectors used for triggering, background rejection, and event characterisation. A detailed description of the apparatus and its performance can be found in Refs.~\cite{Aamodt:2008zz,Abelev:2014ffa}. 

The main detectors used for the track reconstruction and particle identification in the analysis are the Inner Tracking System (ITS)~\cite{Aamodt:2010aa}, the Time Projection Chamber (TPC)~\cite{ALME2010316}, and the Time-Of-Flight (TOF) detector~\cite{Akindinov:2013tea}. They are located inside a large solenoidal magnet, providing a uniform magnetic field of 0.5 T parallel to the LHC beam direction. 
Charged-particle trajectories are reconstructed from their hits in the ITS and the TPC.
The ITS consists of a six-layer silicon detector, which is used for tracking and for the reconstruction of primary and secondary vertices. The TPC is used for track reconstruction and particle identification (PID) via the measurement of the specific energy loss \dEdx. The TOF detector extends at intermediate momenta the PID capabilities of the TPC via the measurement of the time-of-flight of charged particles from the interaction point to the detector.

Event selection and characterisation are performed with the V0 detector~\cite{Abbas:2013taa} and the Zero Degree Calorimeter (ZDC)~\cite{Arnaldi:1999zz}. The V0 detector consists of two scintillator arrays, which cover the full azimuth in the pseudorapidity intervals $-3.7< \eta <-1.7$ (V0C) and $2.8< \eta <5.1$ (V0A), and is used for triggering, event selection, and centrality determination. A minimum-bias interaction trigger was provided by the coincidence of signals in the V0A and the V0C. Furthermore, two separate trigger classes, which consist of an online event selection based on the V0 signal amplitude, were used during the 2018 \PbPb data taking period in order to enrich the sample of central and semicentral collisions.
Events due to the interaction of the beams with residual gas in the vacuum pipe were rejected offline using the V0 and the ZDC timing information. Only events with a primary vertex reconstructed within $\pm10$~cm from the centre of the detector along the beam axis were considered in the analysis.

Collisions were divided into centrality classes, defined in terms of percentiles of the hadronic Pb--Pb cross section, using the sum of the V0 signal amplitudes, as described in details in Ref.~\cite{Acharya:2018hre}. 
 The centrality classes used in the present analysis, together with the corresponding average nuclear overlap function $\av{\TAA}$~\cite{Adam:2015ptt, ALICE-PUBLIC-2018-011}, the number of events ($N_{\rm events}$), and the recorded integrated luminosity $L_{\rm int}$~\cite{Loizides:2017ack} in each class, are summarised in Table~\ref{tab:Nevents}.

	\begin{table}[!t]
	\centering
	\caption{Average nuclear overlap function, number of events, and recorded integrated luminosities for the two centrality classes used in the analysis.}
	\begin{tabular}{cccc}
	\hline
        \noalign{\vskip 0.1cm}
	Centrality class & $\langle\TAA\rangle$ (mb$^{-1}$) & $N_{\rm events}$ & $L_{\rm int} (\mu \mathrm{b}^{-1})$\\[0.1cm]
	\hline
        \noalign{\vskip 0.1cm}
	\phantom{0}0--10\% & $23.26\pm0.17$ & $100 \times 10^6$ & 130.5 $\pm$ 0.5 \\
	30--50\% & $3.92\pm0.06$ & $85 \times 10^6$ & 55.5 $\pm$ 0.2\\
	\hline
	\end{tabular}		
	\label{tab:Nevents}
	\end{table}

\section{Analysis techniques}
\label{sec:analysis}

\subsection{D-meson raw yields}
\label{sec:first_method}
 
The D mesons and their charge conjugates were reconstructed via the following hadronic decay channels: $\DtoKpi$ (with branching ratio, $\mathrm{BR}= 3.950\pm0.031\%$), $\DtoKpipi$ ($\mathrm{BR} = 9.38\pm0.16\%$), and $\DstartoDpi$ ($\mathrm{BR}=67.7\pm0.5\%$)~\cite{Zyla:2020zbs}. 
The analysis was based on the reconstruction of decay vertices displaced from the primary vertex, exploiting the separation of few hundreds $\mu$m due to the weak decays of $\Dzero$ and $\Dplus$ (c$\tau$ of $\sim$ 123 and $\sim$ 312 $\mu$m, respectively~\cite{Zyla:2020zbs}). In the case of the strong decay of the $\Dstar$ meson, the decay topology of the produced $\Dzero$ was exploited. This method allowed the reconstruction of $\Dzero$ candidates for $\pt>1~\gevc$ in both centrality classes and of $\Dplus$ ($\Dstar$) candidates for $\pt>$ 2.5 (3)$~\gevc$ and for $\pt>$ 2 (2)$~\gevc$ in the 0--10\% and 30--50\% centrality classes, respectively.

$\Dzero$ and $\Dplus$ candidates were built using pairs and triplets of tracks with proper charge-sign combination, $|\eta| <0.8$, $\pt >$ 0.4 GeV/$c$, a minimum number of two hits (out of six) in the ITS, with at least one in the two innermost layers, at least 70 out of 159 crossed TPC pad rows, and a fit quality \mbox{$\chi^2/$ndf $<$ 1.25} in the TPC (where ndf is the number of degrees of freedom involved in the track fit procedure).
$\Dstar$ candidates were reconstructed by combining $\Dzero$ candidates with tracks, which were required to have $|\eta|<$ 0.8, $\pt>$ 0.1 $\GeV/c$, and at least three associated hits in the ITS.

These track selection criteria reduce the D-meson acceptance in rapidity, which, depending on $\pt$, varies from $|y|>0.5$ at low $\pt$ and $|y|>$ 0.8 for $\pt>$ 5 GeV/$c$.
Consequently, a $\pt$-dependent fiducial acceptance selection, $|y|<y_{\mathrm{fid}}(\pt)$, was applied, with $y_{\mathrm{fid}}(\pt)$ defined according to a second-order polynomial function increasing from 0.5 to 0.8 in the range $0<\pt<5~\GeV/c$, and fixed to 0.8 for $\pt>5~\GeV/c$.

The D-meson selection strategy is similar to the one used in previous analyses and the variables used to distinguish between signal and background candidates are based on the impact parameters (i.e. the distance of closest approach between the track and the primary vertex in the plane transverse to the beam direction) of the decay particles, the separation of the primary and secondary vertices, and the pointing of the reconstructed D-meson momentum to the primary vertex, as described in Refs.~\cite{Adam:2015jda, Acharya:2018hre}. 
These selection criteria tend to enhance reconstructed feed-down D mesons, originating from beauty-hadron decays, over those promptly produced. An additional selection was therefore applied on the normalised difference between the measured and expected impact parameters of each of the decay particles, which allows for a significant rejection of background candidates and feed-down D mesons, while keeping a large fraction of prompt D mesons as reported in Ref.~\cite{Acharya:2017jgo}. Further reduction of the combinatorial background was obtained by applying PID for charged pions and kaons with the TPC and TOF detectors. A $\pm$3$\sigma$ window around the expected mean values of \dEdx~and time-of-flight  was used for the identification,  where $\sigma$ is  the  resolution  on  these  two  quantities. A tighter PID selection criterion ($\pm2 \sigma$) was used for $\Dplus$ (for $\pt <$ 3 GeV/$c$) and $\Dstar$ candidate daughters in central collisions because of the large background of track triplets.

\begin{figure*}[!t]
\begin{center}
\includegraphics[width=1.\textwidth]{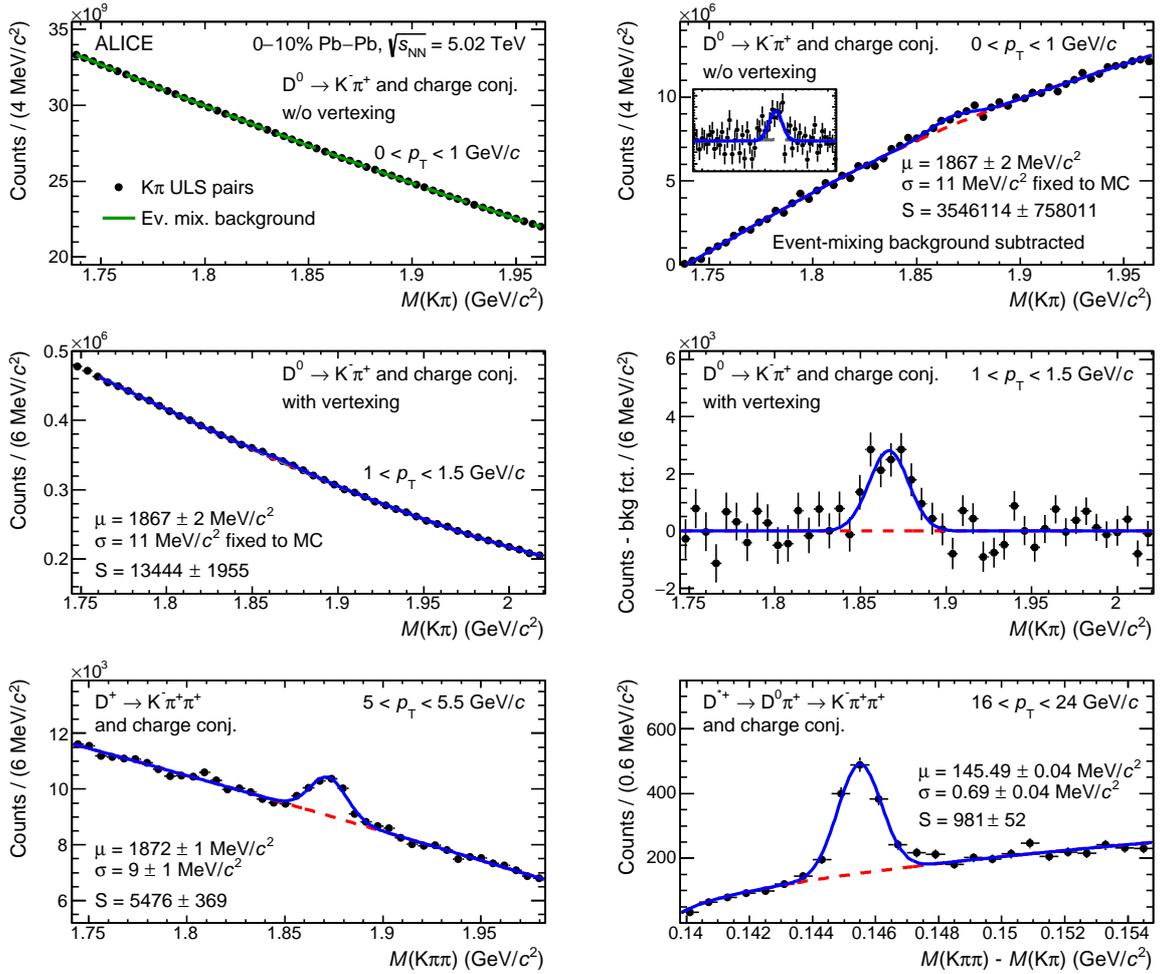}
\caption{Invariant-mass (mass-difference) distribution of $\Dzero$, $\Dplus$, and $\Dstar$ meson candidates in different $\pt$ intervals for the centrality class 0--10$\%$. The fit functions are composed of a Gaussian function for the signal and an additional term for the background, as described in detail in the text. The values for the Gaussian mean $\mu$, width $\sigma$, and raw yield \textit{S} are also reported. Top row: $\Dzero$-meson candidates with $0<\pt<1~\gevc$ without reconstructing the decay vertex. The invariant-mass distributions are shown before (left) and after (right) the subtraction of the combinatorial background estimated from event-mixing.
Middle row: $\Dzero$-meson candidates with $1<\pt<1.5~\gevc$ with reconstruction of the decay vertex before (left) and after (right) background subtraction. The width of the Gaussian in this and in the previous $\pt$ interval is fixed to the value obtained from simulations. Bottom row: $\Dplus$-meson candidates with $5<\pt<5.5~\gevc$ (left) and $\Dstar$-meson candidates with $16<\pt<24~\gevc$ (right).} 
\label{fig:Dmass} 
\end{center}
\end{figure*}

An alternative analysis method, not based on the reconstruction and selection of the displaced decay-vertex topology, was previously developed and applied for the two-body decay of $\Dzero$ mesons in pp and $\pPb$ collisions, in order to extend the measurement down to $\pt= 0$~\cite{Adam:2016ich, Acharya:2019mgn}. This technique, which is used for the first time in the analysis of \PbPb collisions, does not use geometrical selections on the displaced decay vertex (which are not effective at very low $\pt$), but relies mainly on particle identification and on the estimation and subtraction of the combinatorial background. 
The $\Dzero$ candidates were formed combining pairs of tracks with opposite charge sign (unlike sign, ULS), with $|\eta|<$ 0.8 and $\pt>$ 0.4 $\gevc$. Single-track and particle identification selections as well as the fiducial acceptance selection of the reconstructed candidates were performed using the same strategy adopted in the analysis with decay-vertex reconstruction previously described. 
The contribution to the invariant-mass distribution of ULS K$\pi$ pairs due to combinatorial background is estimated with the event-mixing technique. The events are grouped in pools, before the  event-mixing, based on the primary vertex position along \textit{z} and on the event multiplicity, quantified by the number of track segments reconstructed in the two innermost layers of the ITS.  Then, the kaon tracks of a given event are mixed with the pion tracks of other events. The top left panel of Fig.~\ref{fig:Dmass} shows the invariant-mass distribution of ULS K$\pi$ pairs together with that of the background estimated with the event-mixing technique in the interval $0 < \pt < 1~\gevc$. The latter distribution was normalised to match the yield of ULS pairs at one edge of the invariant-mass interval considered for the extraction of the $\Dzero$ raw yield.

The D-meson raw yields, including particles and antiparticles, were obtained from binned maximum-likelihood fits to the $\Dzero$ and $\Dplus$ candidate invariant mass ($M$) distribution and to the mass difference $\Delta M = M (\mathrm{K} \pi \pi) - M(\mathrm{K} \pi)$ distribution for $\Dstar$ candidates. Figure~\ref{fig:Dmass} shows examples of fits for these distributions in the 0--10$\%$ centrality class and for four $\pt$ intervals. In the analysis of the $\Dzero$ meson without decay-vertex reconstruction, the fit is performed after subtracting the estimated background from the ULS K$\pi$ invariant-mass distribution as shown in the top right panel.
 
 The fit function was composed of a Gaussian term to describe the signal and an exponential function for the background for $\Dzero$ and $\Dplus$ candidates at intermediate and high $\pt$ in the analysis with vertex reconstruction. At low $\pt$ ($\pt<3~\gevc$ for $\Dzero$ and $\pt<5~\gevc$ for $\Dplus$), a second-order polynomial function is used to model the background invariant-mass shape in both centrality classes. In the $\Dzero$ analysis without vertex reconstruction, instead, the background was parametrised for $\pt<$ ($>$) 2$~\gevc$ with a fourth-order (third-order) polynomial function.
 The $\Delta M$ distribution of $\Dstar$ candidates was fitted with a Gaussian function for the signal and the following function for the background: $a \sqrt{\Delta M - m_{\pi}}{\rm e}^{b(\Delta M - m_{\pi})}$, where $m_{\pi}$ is the pion mass and $a$ and $b$ are free parameters.
 
 In some cases a $\Dzero$-meson candidate can be also identified as a $\Dzerobar$ meson when the two daughter tracks are compatible with both the kaon and pion hypothesis, leading to an irreducible correlated background. Hence, in the $\Dzero$ meson analyses, an additional term was included in the fit function to take into account this contribution to the invariant mass of signal candidates, which is called reflections. It was parametrised with a double-Gaussian distribution based on detailed Monte Carlo simulations. This contribution ranges between 2\% and 3\% of the raw signal depending on $\pt$~\cite{Acharya:2018hre}.
 Given the critical signal extraction for the  $\Dzero$ meson at low $\pt$ in the analysis with vertex reconstruction, due to the small signal-to-background ratio, the width of the Gaussian for the signal was fixed to the value obtained from the simulation for $\pt < 1.5 ~\gevc$. We verified that the widths from the simulation were consistent with those extracted from the data. For the same reason, in the analysis without vertexing the Gaussian width was fixed to the one from the simulation in the whole $\pt$ range.
 The statistical significance of the signal, defined as $S/\sqrt{S+B}$ where $S$ is the raw signal yield obtained by integrating the Gaussian function and $B$ is the background under the peak (within 3$\sigma$), varies from 4 to 60, depending on the D-meson species, the $\pt$ interval, and the centrality class. In the $\Dzero$ analysis without decay-vertex reconstruction, the $S/B$ in the $\pt$ range 0--1 $\gevc$ is approximately $\sim$~8.5$\times$10$^{-6}$ before the mixed-event background subtraction.

\subsection{Yield corrections and beauty feed-down subtraction}
\label{sec:corrected_yields}
 The D-meson raw yields were corrected to obtain the $\pt$-differential yields of prompt D-mesons according to:
 \begin{equation}
  \frac{\mathrm{d^2}N}{\mathrm{d}y~\mathrm{d}p_{\mathrm{T}}} = \frac{1}{2}\frac{1}{\Delta p_{\mathrm{T}}} \frac{f_{\mathrm{prompt}}(p_{\mathrm{T}})\times N^{\mathrm{D+\overline D}}_{\mathrm{raw}}(p_{\mathrm{T}})\Big\rvert_{|y| < y_{\mathrm{fid}}}}{\alpha_{y}(p_{\mathrm{T}})\times(\mathrm{Acc}\times\epsilon)_{\mathrm{prompt}}(p_{\mathrm{T}})\times \mathrm{BR}\times N_{\mathrm{events}} } .
  \label{eq:dNdpt}
 \end{equation}
 The raw yield values $N^{\rm D+\overline D,raw}$, which contain the contribution of feed-down from beauty-hadron decays, were multiplied by the fraction of promptly produced D mesons $f_{\rm{prompt}}$ and divided by a factor of two to obtain the charged-averaged yields. Furthermore, they were divided by the product of prompt D-meson acceptance and efficiency ($\mathrm{Acc}\times\epsilon)_\mathrm{prompt}$, the branching ratio BR of the decay channel, the width of the $\pt$ interval ($\Delta \pt$), and by the number of events  $N_{\mathrm{events}}$.
 The factor $\alpha_{y}(p_{\mathrm{T}})$ normalises the corrected yields measured in $\Delta y =$ 2$y_{\mathrm{fid}}(p_{\mathrm{T}})$ to one unit of rapidity.
It was computed as the ratio between the generated D-meson yield in $|y|<y_{\mathrm{fid}}(p_{\mathrm{T}})$ and that in $|y|<0.5$ using a data-driven $\pt$ shape and a rapidity distribution from fixed order plus next-to-leading logarithms (FONLL)~\cite{Cacciari:1998it,Cacciari:2001td} perturbative QCD calculations, after verifying that the D-meson rapidity distributions in FONLL are consistent with those from PYTHIA 8~\cite{Sjostrand:2014zea}.

The ($\mathrm{Acc}\times\epsilon$) correction was obtained separately for prompt and feed-down D mesons using simulations with the GEANT3 transport package~\cite{Brun:1994aa}, including a detailed description of the detector geometry and response as well as of the LHC beam conditions. The \PbPb collisions at $\sqrtsNN = 5.02~\TeV$ were produced with the HIJING v1.36~\cite{Wang:1991hta} event generator and D-meson signals were added by injecting ${\rm c \overline{c}}$ or ${\rm b \overline{b}}$ pairs generated with the PYTHIA v8.2.43 event generator with Monash tune~\cite{Skands:2014pea}. 
The D-meson $\pt$ distributions from the simulations were reweighted in order to use realistic momentum distributions in the determination of the acceptance and the efficiency, which depend on $\pt$. The weights were defined with an iterative procedure to match the $\pt$ dependence measured for $\Dzero$ mesons in the intervals used in the analysis, for both centrality classes and in the momentum interval 1$< \pt <$ 50$~\gevc$.

\begin{figure}[!t]
\begin{center}
\includegraphics[width=0.4\textwidth]{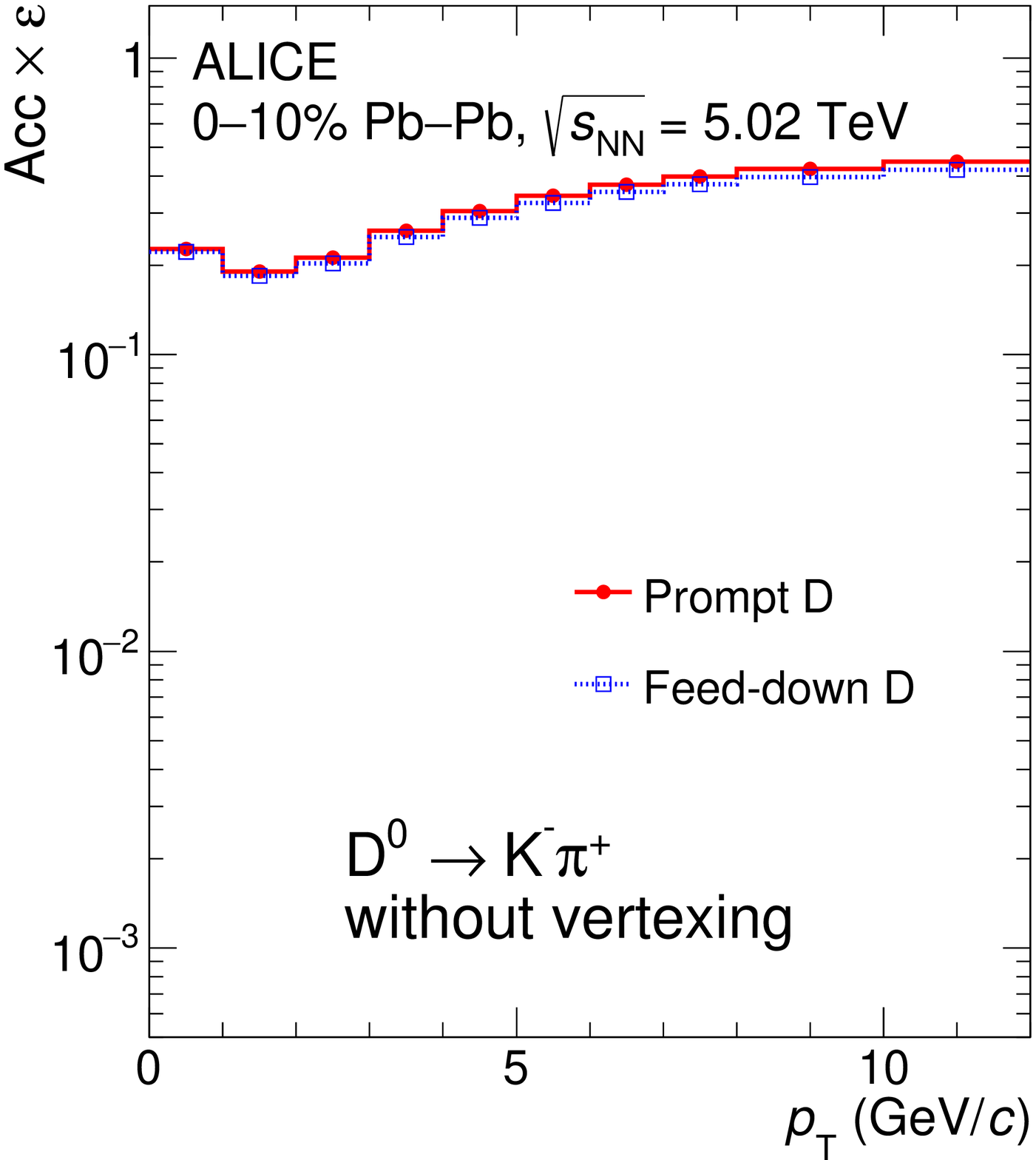}
\includegraphics[width=0.4\textwidth]{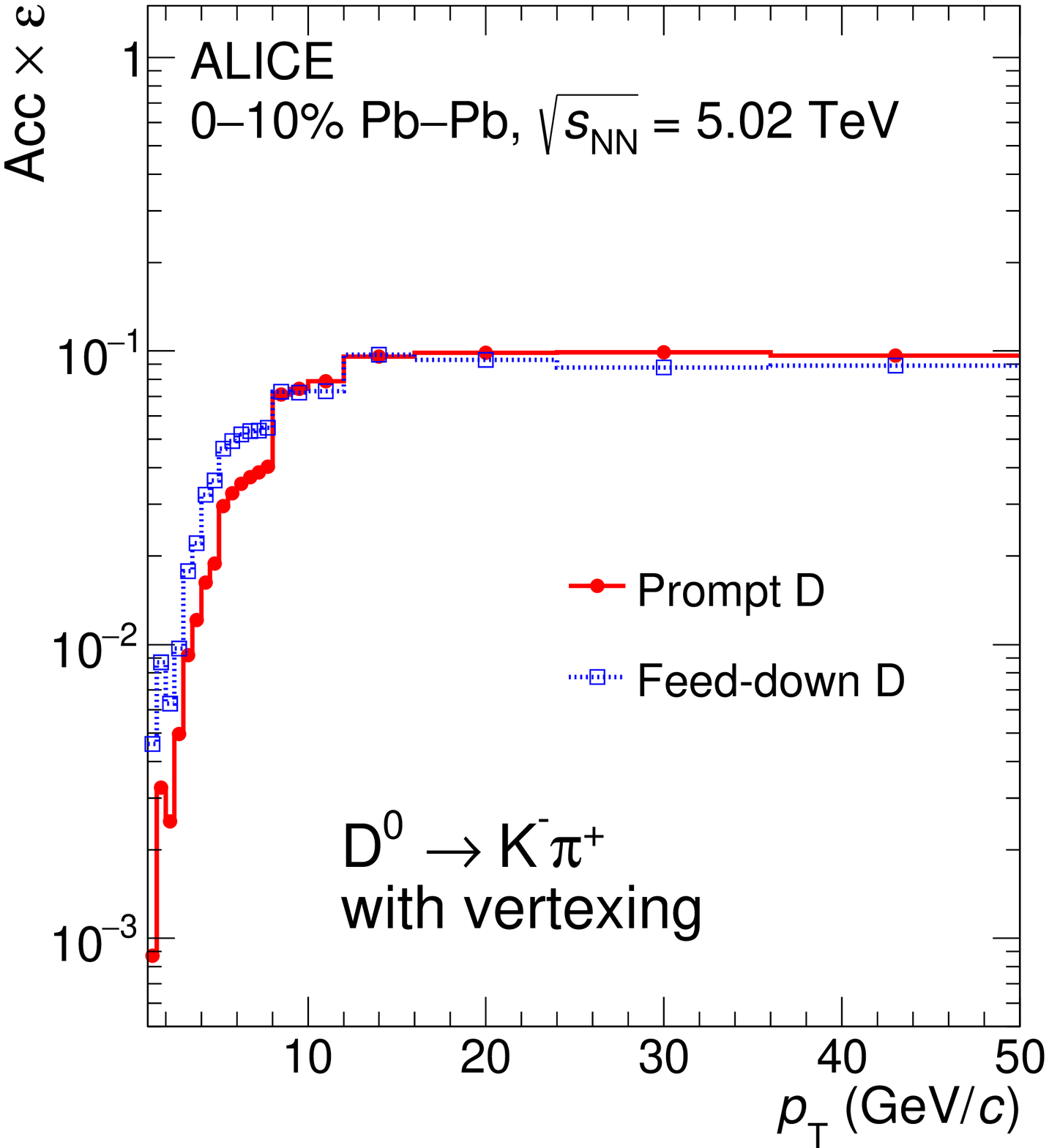}
\includegraphics[width=0.4\textwidth]{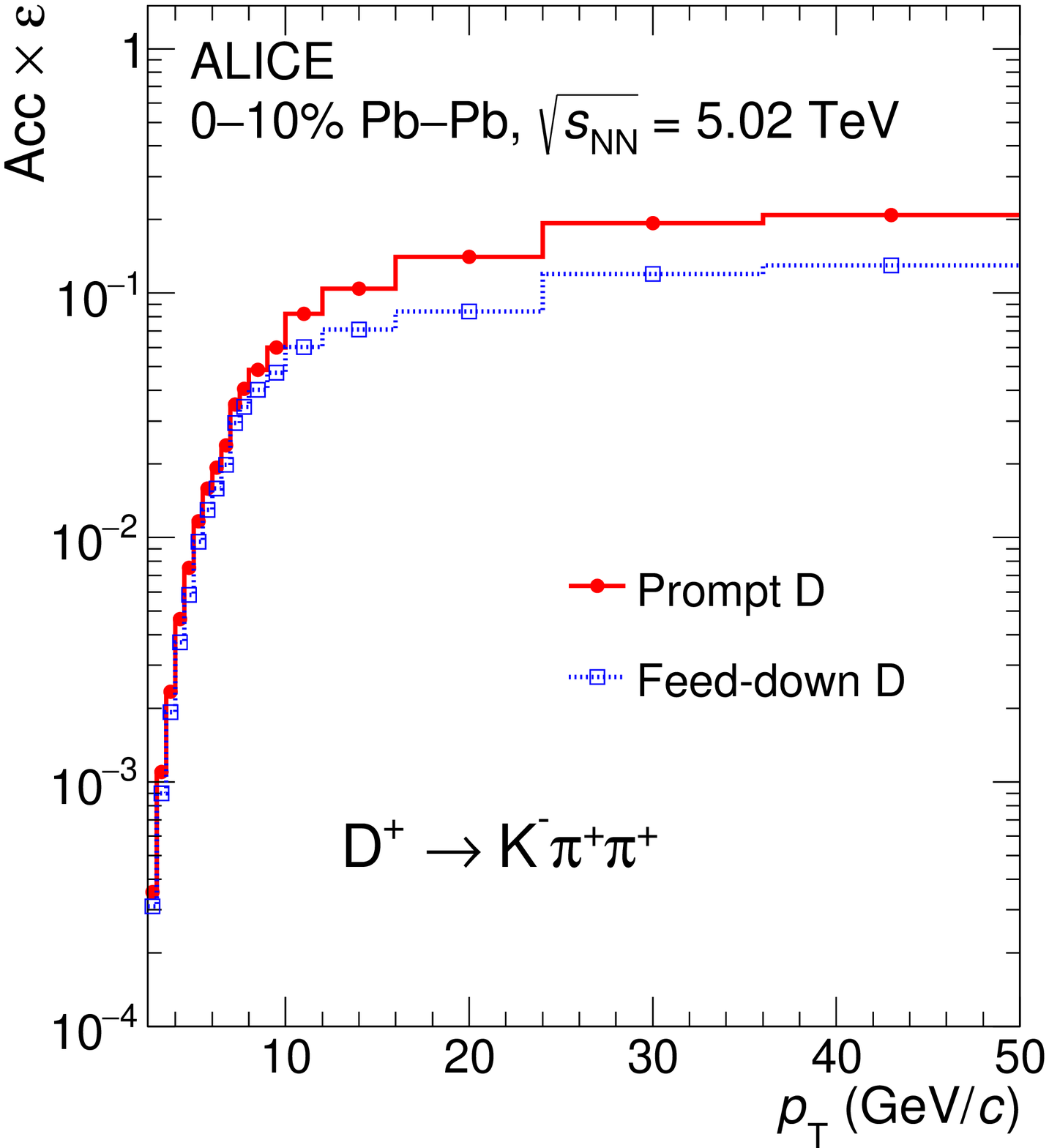}
\includegraphics[width=0.4\textwidth]{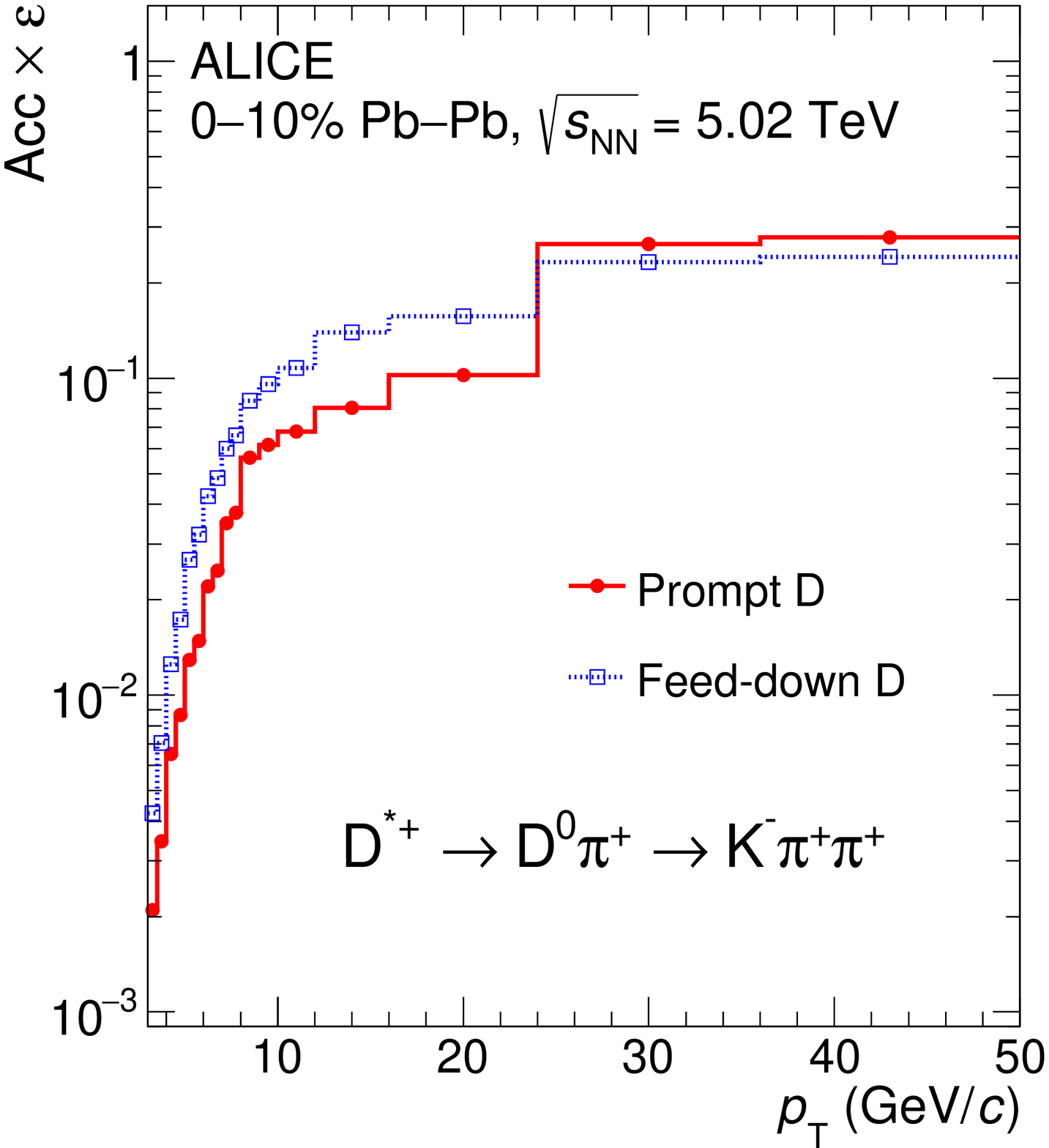}
\caption{Product of acceptance and efficiency ($\mathrm{Acc}\times\epsilon$) as a function of $\pt$ for prompt (red circles) and feed-down (blue squares) D mesons in \PbPb collisions for the 0--10\% centrality class.  
} 
\label{fig:Deff}
\end{center}
\end{figure}
Figure~\ref{fig:Deff} shows the ($\mathrm{Acc}\times\epsilon$) for prompt and feed-down $\Dzero$ (top panels), $\Dplus$ (bottom left panel), and $\Dstar$ (bottom right panel) mesons with $|y|<y_{\mathrm{fid}}$ as a function of $\pt$ in the 0--10\% centrality class.
For the analysis that does not exploit the selections on the $\Dzero$-meson decay vertex (top left panel), the efficiency is higher by a factor of about 100 (5) at low (high) $\pt$ as compared to the analysis with decay-vertex reconstruction (top right panel).
The $({\rm Acc}\times\epsilon)$  is almost independent of $\pt$ (the mild increase with increasing $\pt$ is mainly determined by the geometrical acceptance of the detector) and is the same for prompt and feed-down $\Dzero$, as expected when no selection is made on the displacement of the $\Dzero$-meson decay vertex from the interaction point.

The efficiencies for the analysis with geometrical selections on the displaced decay-vertex topology range from about 10$^{-3}$ at low $\pt$ to 0.1--0.3 at high $\pt$, depending on the D-meson species. The decreasing efficiency with decreasing $\pt$ is due to the fact that more stringent selection criteria are needed at low $\pt$ because of the larger background and the smaller average displacement of the decay vertex from the interaction point. The trend of the efficiency with $\pt$ is not completely smooth, especially for $\Dzero$ and $\Dstar$, reflecting the $\pt$ intervals in which the optimisation of the selection criteria was performed. The ($\mathrm{Acc}\times\epsilon$) is higher for semicentral collisions than for central collisions, since less stringent selections are applied because of the lower combinatorial background.
The difference between the efficiencies for prompt and feed-down D mesons is due to the geometrical selections applied since the latter are more displaced from the primary vertex due to large B-meson lifetime ($\tau \approx$ 500 $\mu$m) and are therefore more efficiently selected by the majority of the selection criteria applied in the analysis. 
However, the requirement on the difference between measured and expected decay track impact parameter rejects efficiently D mesons coming from beauty-hadron decays. This is particularly effective for $\Dplus$ mesons (Fig.~\ref{fig:Deff} bottom left panel) whose feed-down efficiencies become lower than the prompt efficiencies after a selection on the aforementioned variable is applied.

The fraction of selected prompt D mesons $f_{\rm prompt}$ was obtained with the procedure introduced in Refs.~\cite{ALICE:2012ab, Acharya:2018hre} to account for the contribution of D mesons from beauty-hadron decays in the measured raw yield. The $f_{\rm prompt}$ factor was estimated in each $\pt$ interval using perturbative QCD calculations for the cross section of feed-down D mesons, efficiencies from the simulations, and a hypothesis on the $\RAA$ of feed-down D mesons. In detail, the expression for $f_{\rm prompt}$ is:
\begin{equation}
  \label{eq:fcNbMethod}
\begin{split}
f_{\rm prompt} &= 1-\frac{ N^{\rm D+\overline D\,\textnormal{feed-down}}_{\rm raw}} { N^{\rm D+\overline D}_{\rm raw} }\\
 &= 1 -    \RAA^{\textnormal{feed-down}} \times    \langle \TAA \rangle 
	 \times \left( \frac{{\rm d} \sigma}{{\rm d}\pt }
         \right)^{\textnormal{FONLL,\,PYTHIA 8}} _{{\textnormal{
               feed-down},\,|y|<0.5}} \times
         \frac{2\Delta\pt\times\alpha_y\times({\rm
             Acc}\times\epsilon)_{\textnormal{feed-down}}\times {\rm
             BR} \times N_{\rm events}  }{N^{\rm D+\overline D}_{\rm raw}  } \, ,
\end{split}
\end{equation}
where $N^{\rm D+\overline D}_{\rm raw}$ is the measured raw yield
and $N^{{\rm D+\overline D}\,\textnormal{feed-down}}_{\rm raw}$ is the estimated raw yield of $\rm D$ mesons from beauty-hadron decays. The beauty-hadron production cross section in pp collisions at $\sqrt{s}=5.02~\TeV$, estimated with FONLL calculations~\cite{Cacciari:2012ny}, was folded with the beauty-hadron${\rightarrow \rm D}+X$ decay kinematics from PYTHIA~8 and multiplied by the $({\rm Acc}\times\varepsilon)$ of feed-down $\rm D$ mesons, by the $\langle \TAA \rangle$ of the corresponding centrality class, and by the other factors introduced in Eq.~\ref{eq:dNdpt}. Finally, the nuclear modification factor of D mesons from beauty-hadron decays ($\RAA^{\textnormal{feed-down}}$) was accounted for. 

The comparison of the $\RAA$ of prompt D mesons ($\RAA^{\rm prompt}$) at $\sqrtsNN = 5.02~\TeV$~\cite{Acharya:2018hre} with that of $\rm J/\psi$ from B-meson decays at the same energy measured by the CMS~\cite{Sirunyan:2017isk} and the ATLAS~\cite{Aaboud:2018quy} collaborations indicates that prompt charmed hadrons are more suppressed than non-prompt charmed hadrons. The difference between the $\Raa$ values is about a factor two in central collisions at a $\pt$ value of about $10~\GeV/c$~\cite{Acharya:2018hre} and it is described by model calculations with parton-mass-dependent energy loss. Therefore, for the centrality classes 0--10\% and 30--50\%, the value $\RAA^{\textnormal{feed-down}}=2\times\RAA^{\rm prompt}$ was assumed to compute the $f_{\rm prompt}$ factor for D mesons with $3<\pt<24~\GeV/c$.
To estimate a systematic uncertainty, this hypothesis was varied in the range $1<\RAA^{\textnormal{feed-down}}/\RAA^{\rm prompt}<3$ considering the data uncertainties and model variations. 
For $\pt<3~\GeV/c$ and $24<\pt<50~\GeV/c$, since model calculations predict a lower difference between $\RAA^{\rm prompt}$ and $\RAA^{\textnormal{feed-down}}$~\cite{Djordjevic:2015hra,He:2014cla}, the hypothesis $\RAA^{\textnormal{feed-down}}=1.5\times\RAA^{\rm prompt}$ was used, with a variation in the range $1<\RAA^{\textnormal{feed-down}}/\RAA^{\rm prompt}<2$ for estimating the systematic uncertainty.

The resulting values of the $f_{\rm prompt}$ factor, for the central hypotheses on $\RAA^{\textnormal{feed-down}}/\RAA^{\rm prompt}$, range from about 0.80 to 0.95, depending on the D-meson species, centrality class, and $\pt$ interval. The values of $f_{\rm prompt}$ for the $\Dzero$ analysis without decay-vertex reconstruction are larger compared to those from the analysis with geometrical selections since the feed-down component is not enhanced by the topological selection criteria.
The systematic uncertainties obtained from the variation of the hypotheses are discussed in Section~\ref{sec:systematics}.

\section{Systematic uncertainties}
\label{sec:systematics}

The systematic uncertainties on the corrected yields of $\Dzero$, $\Dplus$, and $\Dstar$ mesons were estimated as a function of $\pt$ for the 0--10\% and 30--50\% centrality classes considering the following sources: (i) extraction of the raw yield from the invariant-mass distributions; (ii) reconstruction efficiency of the decay-particle tracks; (iii) D-meson selection efficiency; (iv) PID efficiency; (v) generated D-meson $\pt$ and rapidity shape; (vi) subtraction of the contribution originating from beauty-hadron decays.
In addition, a global normalisation uncertainty due to the branching-ratio uncertainty and the centrality interval determination was considered.
The estimated values of the systematic uncertainties are summarised in Table~\ref{tab:syst} for the three D-meson species in representative $\pt$ intervals.

\begin{table}[!tb]
\begin{center}
\caption{Relative systematic uncertainties on prompt D-meson yields in Pb--Pb collisions at $\sqrtsNN = 5.02~\TeV$ for representative $\pt$ intervals. The first $\pt$ interval of the $\Dzero$ corresponds to the analysis without decay vertex reconstruction. For the uncertainties on the correction factors, values below 0.5\% are considered negligible.}
\begin{tabular}{|c|l|ccc|cc|cc|}
\hline
\multicolumn{2}{|r|}{\rule{0pt}{2.6ex} Particle} & \multicolumn{3}{c|}{$\Dzero$} & \multicolumn{2}{c|}{$\Dplus$}  & \multicolumn{2}{c|}{$\Dstar$} \\[0.5ex]
\hline
\multicolumn{2}{|r|}{\rule{0pt}{2.6ex} $\pt$ interval ($\gevc$)} & 0--1  & 1.5--2  & 7.5--8        & 3--3.5  & 7.5--8                 & 3--3.5  & 7.5--8 \\[0.5ex]
\hline
\hline
\multirow{9}{*}{\parbox{2 cm}{\centering 0--10\%\\ centrality}} 
& Raw-yield extraction \rule{0pt}{2.6ex}            & 9\%  & 7\%  & 2\%             & 5\%  & 5\%                     & 7\%  & 3\%\\
& Correction factor & & & & & & &\\
& $\qquad$ Tracking efficiency   & 8\%  & 9\%  & 9\%         & 14\%  & 15.5\%               & 13\%  & 9\%\\
& $\qquad$ Selection efficiency  & negl.    & 8\%  & 2\%             & 5\%  & 5\%                     & 13\%  & 4\%\\
& $\qquad$ PID efficiency        & negl.    & negl.    & negl.             & negl.  & negl.                       & 1\%  & 1\%\\
& $\qquad$ MC $\pt$ shape        & negl.    & 0.5\%  & negl.               & negl.  & negl.                       & 0.5\%  & negl. \\[1ex]
& Feed-down from beauty           & $^{+2.7}_{-2.9}\%$ & $^{+7.6}_{-8.3}\%$ & $^{+11.6}_{-12}\%$    & $^{+3.4}_{-3.2}\%$ & $^{+7.1}_{-6.6}\%$    & $^{+8.4}_{-8.1}\%$ & $^{+10.3}_{-10.3}\%$\\[1ex]	
\cline{3-9} 
& Branching ratio \rule{0pt}{2.6ex}      & \multicolumn{3}{c|}{0.8\%}      &\multicolumn{2}{c|}{1.7\%}        &\multicolumn{2}{c|}{1.1\%}\\
\cline{3-9} 
& Centrality limits \rule{0pt}{2.6ex}    & \multicolumn{7}{c|}{<0.1\%}\\
\hline
\hline
\multirow{9}{*}{\parbox{2 cm}{\centering 30--50\%\\ centrality}} 
& Raw-yield extraction \rule{0pt}{2.6ex}           & 11\%  & 4\%  & 2\%             & 4\%  & 3\%                     & 5\%  & 4\%\\
& Correction factor & & & & & & &\\
& $\qquad$ Tracking efficiency   & 7\%  & 7.5\%  & 7\%           & 11\%  & 11\%               & 10\%  & 9\%\\
& $\qquad$ Selection efficiency  & negl.    & 8\%  & 2\%             & 4\%  & 3\%                     & 9\%  & 4\%\\
& $\qquad$ PID efficiency        & negl.    & negl.    & negl.           & negl.    & negl.                   & negl.    & negl. \\
& $\qquad$ MC $\pt$ shape        & negl.    & 1\%  & negl.               & negl.  & negl.                     & negl.    & negl. \\[1ex]
& Feed-down from beauty       & $^{+2.7}_{-2.9}\%$ & $^{+9.6}_{-10.4}\%$ & $^{+11.4}_{-11.7}\%$   & $^{+3.4}_{-3.2}\%$ & $^{+6.6}_{-6.3}\%$   & $^{+8.5}_{-8.2}\%$ & $^{+9.9}_{-9.9}\%$\\[1ex]	
\cline{3-9} 
& Branching ratio \rule{0pt}{2.6ex}      & \multicolumn{3}{c|}{0.8\%}      &\multicolumn{2}{c|}{1.7\%}        &\multicolumn{2}{c|}{1.1\%}\\
\cline{3-9} 
& Centrality limits \rule{0pt}{2.6ex}    & \multicolumn{7}{c|}{2\%}\\
\hline
\end{tabular}
\label{tab:syst}
\end{center}
\end{table}

The systematic uncertainty on the D-meson raw yield was estimated in each $\pt$ interval by varying the invariant-mass interval considered in the fit and the functional form of the background fit function.
The sensitivity to the line shape of the signal was tested by considering the raw yield values obtained by counting the candidates in the invariant-mass region of the signal peak after subtracting the background estimated from the side bands. 
In the case of the $\Dzero$-meson analysis without decay-vertex reconstruction, an additional contribution related to the line shape of the signal was estimated by varying the width of the Gaussian function by $\pm 10\%$ with respect to the Monte Carlo expectation, based on the deviations between the Gaussian width values observed in data and simulations for the analysis with decay-vertex reconstruction.
The systematic uncertainty was defined as the RMS of the distribution of the signal yields obtained from all these variations.
In the case of $\Dzero$ mesons, an additional contribution due to the modelling of the reflection contribution in the fit was estimated by varying (by $\pm 20\%$) the ratio of the integral of the reflections to the integral of the signal and the shape of the templates used in the invariant-mass fits.
The assigned uncertainty ranges from 2\% to 11\% depending on the D-meson species and $\pt$ interval, being on average smaller in the 30--50\% centrality class due to the larger signal-to-background ratio in this class as compared to central collisions.

The systematic uncertainty on the track reconstruction efficiency has two contributions.
The first one was estimated by varying the track quality selection criteria.
The second contribution originates from possible differences in the probability to match the TPC tracks to the ITS hits in data and in simulations.
It was estimated by comparing the matching efficiency in data and simulations after weighting the relative abundances of primary and secondary particles in the simulation to match those observed in data. The uncertainty was estimated for the single track and propagated to the reconstructed D mesons using the decay kinematics.
The estimated uncertainty is the quadratic sum of the two contributions and it depends on the D-meson species and $\pt$, ranging from 3\% to 10.5\% for the two-body decay of $\Dzero$ mesons and from 5.5\% to 15.5\% for the three-body decays of $\Dplus$ and $\Dstar$ mesons.

The uncertainty on the D-meson selection efficiency originates from imperfections in the description of the D-meson kinematic properties and of the detector resolution and alignment in the simulation.
For the analyses with decay-vertex reconstruction, the systematic uncertainty was estimated by comparing the corrected yields obtained by repeating the analysis with different sets of selection criteria resulting in a significant modification of the raw yields, signal-to-background ratios, and efficiencies.
The estimated uncertainty depends on the D-meson species and $\pt$ interval, being larger at low $\pt$ and for central collisions, where more stringent selections are used to obtain a good statistical significance of the signal.
The values obtained in these analyses range from 2\% (3\%) to 8\% (13\%) for the two-(three-) body decay channel.
In the case of the $\Dzero$-meson analysis without decay-vertex reconstruction, no geometrical selections on the displaced decay-vertex topology are applied, and the efficiencies are higher than those of the analysis with decay-vertex reconstruction and almost independent of $\pt$. The stability of the corrected yield was tested against variations of the single-track $\pt$ selection and no systematic effect was observed.

The uncertainty on the PID selection efficiency was estimated by repeating the analyses with decay-vertex reconstruction without applying the PID selections.
The resulting corrected yields were found to be compatible with those obtained with the PID selection and therefore no systematic uncertainty was assigned.
For  the $\Dzero$-meson analysis without decay-vertex reconstruction, the analysis without applying PID selections could not be performed due to the insufficient statistical significance of the signal.
More stringent PID criteria (at 2$\sigma$ level on TPC, or TOF or both) were tested and compatible values for the corrected yields were obtained.
Based on this result and on the fact that the PID selections are the same as used in the analysis with decay-vertex reconstruction, no uncertainty due to PID was assigned.
In the case of the $\Dstar$ analysis, more stringent PID selection criteria were used in the 0--10\% centrality class.
A systematic uncertainty of 1\% was estimated by comparing the pion and kaon PID selection efficiencies in the data and in the simulation and combining the observed differences using the $\Dstar$-meson decay kinematics. For this study, pure pion samples were selected from strange-hadron decays, while pure kaon samples in the TPC (TOF) were obtained using a tight PID selection in the TOF (TPC).

An additional contribution to the systematic uncertainty on the efficiency originates from a possible difference between the real and simulated D-meson $\pt$ and rapidity distributions.
The effect of the $\pt$ shape was estimated by calculating the efficiency using alternative D-meson transverse momentum shapes via a reweighting technique.
In particular, the $\pt$ distributions from FONLL calculations with and without hot-medium effects parametrised based on the $\RAA$ in central collisions from different models were considered.
For the analyses with decay-vertex reconstruction, the resulting uncertainty, which also includes the effect of the $\pt$ dependence of the nuclear modification factor, was estimated to be negligible for $\pt> 5~\gevc$ and to increase to 1--1.5\% in the lowest $\pt$ intervals considered in the analysis, where the efficiency varies steeply with $\pt$.
Instead, no sensitivity to the  generated $\Dzero$ $\pt$ shape was observed in the results of the $\Dzero$-meson analysis without decay-vertex reconstruction.
The simulated rapidity shape of D mesons affects the $\mathrm{Acc}\times\epsilon$ and the $\alpha_{y}$ factors in the calculation of the cross section. It was verified that the D-meson rapidity distributions in the PYTHIA 8 simulations and FONLL calculations are similar, resulting in a negligible effect on the $\mathrm{Acc}\times\epsilon$ and $\alpha_{y}$ correction factors. For the latter factor, an extreme assumption of a flat rapidity shape was tested and the difference with respect to the FONLL case was found to be smaller than 1\% for $\pt <$ 10~$\gevc$ and to be smaller than 2\% at higher $\pt$. Considering that at high $\pt$ the assumption of a flat rapidity shape is an extreme variation, the effect of the generated $\dNdy$ shape was considered to be negligible and no systematic uncertainty was assigned.

The systematic uncertainty on the $f_{\rm prompt}$ correction factor was estimated by varying (i) the FONLL parameters (b-quark mass and factorisation and renormalisation scales, according to the prescription in Ref.~\cite{Cacciari:2012ny}) in the calculation of the $\pt$-differential production cross section of feed-down D mesons, and (ii) the ratio between the feed-down and prompt D-meson $\RAA$, as described above.
The resulting uncertainty ranges between 2\% and 14\%, depending on the D-meson species and $\pt$ and centrality interval.

The normalisation uncertainty due to the centrality interval definition was estimated from the variation of raw yield observed when varying the limits of the centrality classes to account for the uncertainty on the fraction of the hadronic cross section used in the Glauber fit to determine the centrality percentiles (see Ref.~\cite{Adam:2015sza} for details).

In the calculation of the nuclear modification factor, the systematic uncertainties on the D-meson yield in Pb--Pb collisions and on the pp reference cross section were propagated as uncorrelated, except for the uncertainty on the BR, which cancels out in the ratio, and the contribution to the feed-down uncertainty originating from the variation of the parameters of the FONLL calculation, which was considered to be fully correlated between the Pb--Pb and pp measurements.
In particular, the contributions of the uncertainties on (i) the normalisation of the pp cross section (due to the luminosity determination), (ii) the centrality limits of the Pb--Pb samples, and (iii) the $\av{\TAA}$ estimated with the Glauber model are common to all the $\pt$ intervals and therefore they constitute a normalisation uncertainty on the $\RAA$, which is shown separately from the other sources when displaying the results.

\section{Results}
\label{sec:results}

\subsection{Transverse-momentum-differential yields}
\begin{figure*}[!t]
\begin{center}
\includegraphics[width=0.495\textwidth]{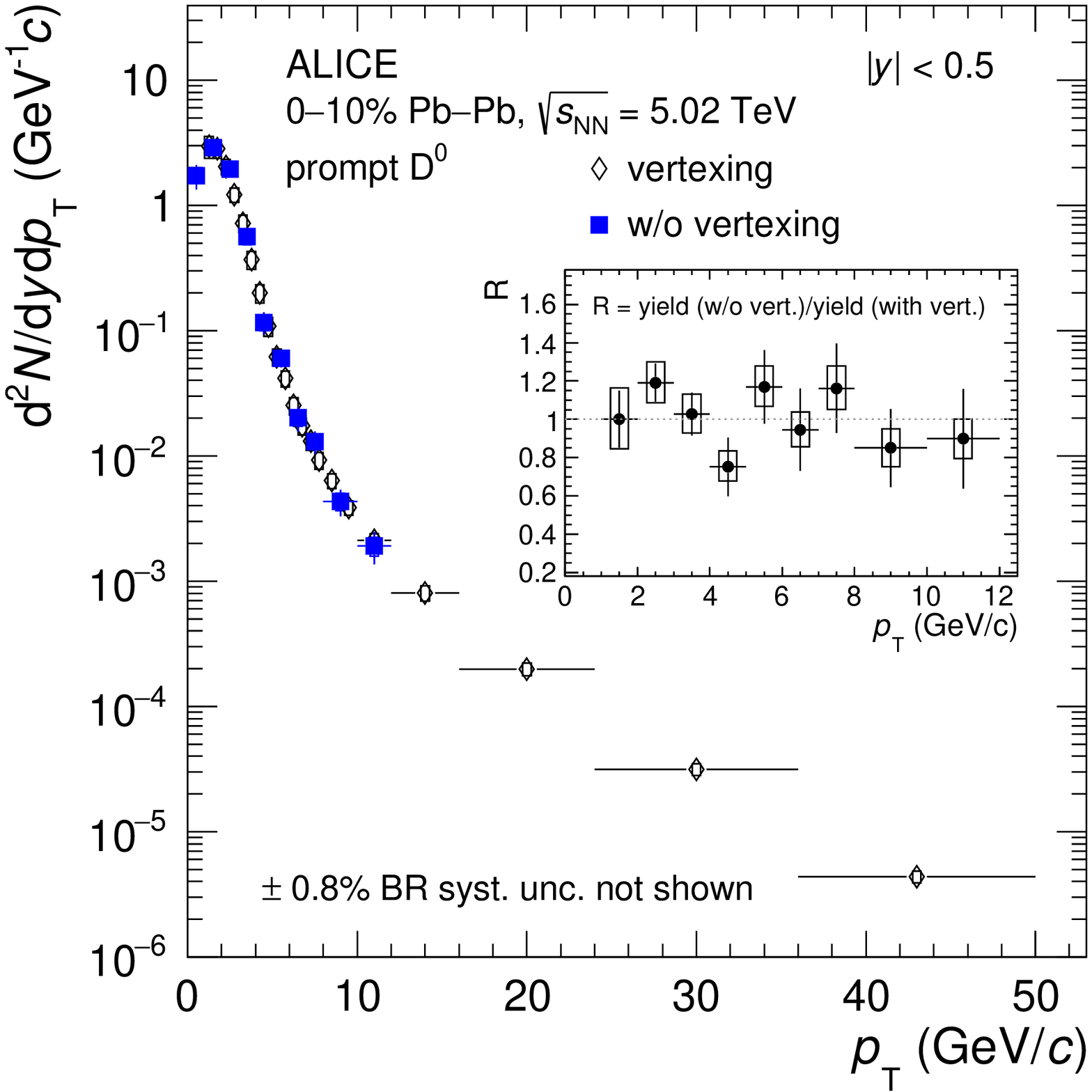}
\includegraphics[width=0.495\textwidth]{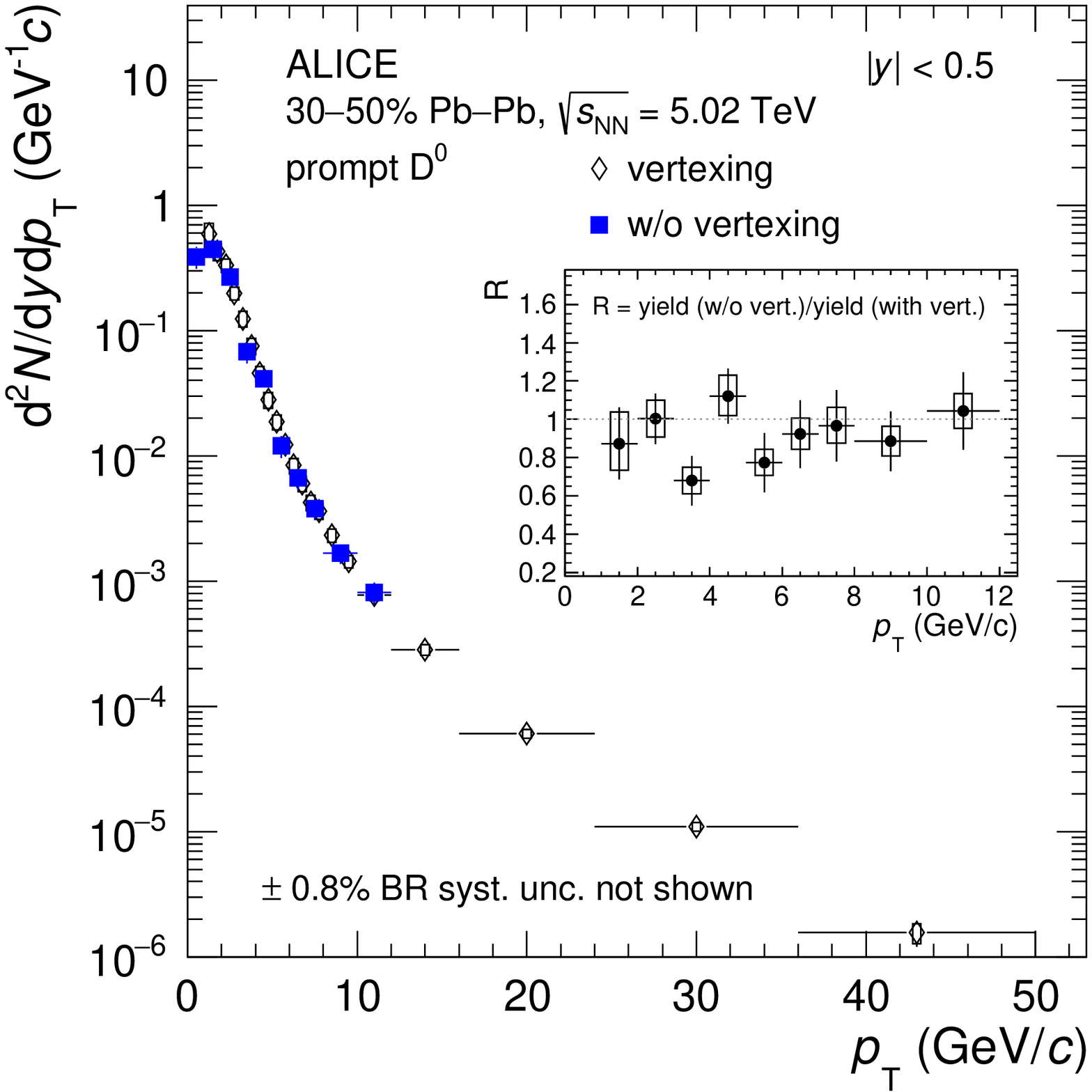}
\caption{Transverse momentum distributions $\dNdydpt$ of prompt $\Dzero$ mesons from the analysis with (open marker) and without (full marker) decay-vertex reconstruction in the 0--10\% and 30--50\% centrality classes in \PbPb collisions at $\sqrtsNN = 5.02~\TeV$. Statistical uncertainties (bars) and systematic uncertainties (boxes) are shown.} 
\label{fig:Dzero_spectra} 
\end{center}
\end{figure*}

In this section, the results on the $\pt$-differential ($\dNdydpt$) and $\pt$-integrated ($\dNdy$) yields of prompt $\Dzero$, $\Dplus$, and $\Dstar$ mesons at midrapidity in \PbPb collisions at $\sqrtsNN = 5.02~\TeV$ in the 0--10\% and 30--50\% centrality classes are presented. 
In the case of the $\Dzero$ mesons, two results for the $\dNdydpt$ were obtained from the analyses with and without selections on the decay-vertex topology. They are compared in Fig.~\ref{fig:Dzero_spectra}, with the inset showing their ratio in the common $\pt$ range. 
In the calculation of the ratio, the results in the narrower $\pt$ intervals of the analysis with decay-vertex reconstruction were merged together to match the binning of the analysis without vertexing, and the systematic uncertainties were propagated treating the raw yield extraction as uncorrelated between the two analyses, and all the other sources of uncertainty as correlated.
The vertical error bars represent the statistical uncertainties and the systematic uncertainties are depicted as boxes except for the BR uncertainty, which is reported separately. The symbols, representing the data points, are positioned horizontally at the center of each interval and the horizontal bars represent the width of the $\pt$ interval. This convention is adopted in all the figures shown in this section.

The two results for the $\pt$-differential yield of prompt $\Dzero$ mesons are found to be consistent within statistical uncertainties, which are independent between the two analysis techniques because of the largely different signal-to-background ratios and efficiencies.
The usage of these two techniques allows the measurement of the $\Dzero$ $\pt$-differential yields in a wide transverse momentum range extending down to $\pt =$ 0 for the first time in \PbPb collisions.
The most precise measurement of the prompt $\Dzero$-meson $\pt$ spectrum, which will be shown and used for comparisons throughout the paper, is obtained by using the results of the analysis without decay-vertex reconstruction in the $\pt$ interval $0<\pt<1~\gevc$ and those from the analysis with decay-vertex reconstruction for $\pt>1~\gevc$. 

\begin{figure*}[!t]
\begin{center}
\includegraphics[width=1.\textwidth]{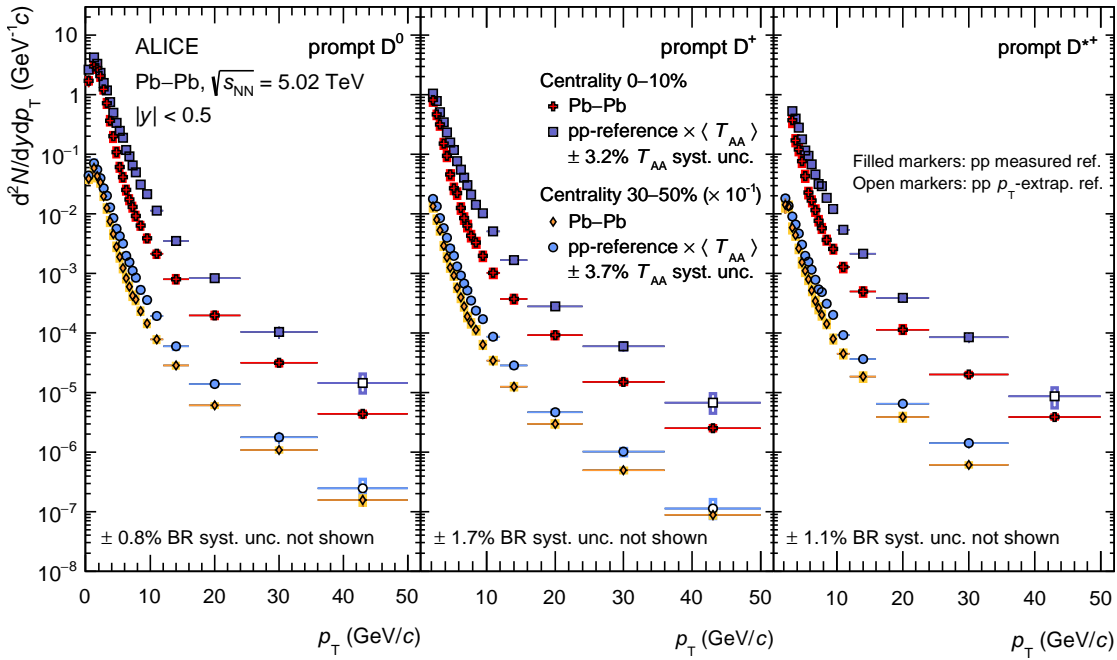}
\caption{Transverse momentum distributions $\dNdydpt$ of prompt $\Dzero$ (left), $\Dplus$ (middle), and $\Dstar$ (right) mesons in the 0--10\% (cross) and 30--50\% (diamond) centrality classes in \PbPb collisions at $\sqrtsNN = 5.02~\TeV$. The reference \pp~distributions multiplied by $\langle\TAA\rangle$ are shown as well. Statistical uncertainties (bars) and systematic uncertainties (boxes) are shown. The uncertainties on the BRs are quoted separately and the horizontal bars represent bin widths. The spectra in the 30--50\% centrality class are scaled by the factor reported in the legend for visibility. Filled and empty markers of the pp reference indicate the measured~\cite{Acharya:2019mgn,Acharya:2021cqv} and $\pt$-extrapolated values, respectively, of the $\pt$-spectrum.} 
\label{fig:Dspectra} 
\end{center}
\end{figure*}

The $\pt$-differential yields $\dNdydpt$ of prompt $\Dzero$, $\Dplus$, and $\Dstar$ mesons are shown in Fig.~\ref{fig:Dspectra} for the 0--10\% and 30--50\% centrality classes. 
They are compared with the reference yields from pp collisions, which are computed as $\langle\TAA\rangle\times\dsigmadpt$, where $\dsigmadpt$ is the D-meson $\pt$-differential cross section measured in pp collisions at $\sqrts = 5.02~\TeV$~\cite{Acharya:2019mgn,Acharya:2021cqv}, and $\langle\TAA\rangle$ is the average nuclear overlap function~\cite{ALICE-PUBLIC-2018-011}. The D-meson production cross sections in pp collisions are measured up to 36~$\gevc$ and they are extrapolated with FONLL towards higher $\pt$, with the method described in Refs.~\cite{Acharya:2018hre,Adam:2015sza}. The spectra in the 30--50\% centrality class are scaled by the factor reported in the legend for visibility. A clear suppression of the production yield for $\pt > 3~\gevc$ is visible in \PbPb collisions and it is stronger in central than in semicentral collisions. 

\begin{figure}[!t]
\begin{center}
\includegraphics[width=1.05\textwidth]{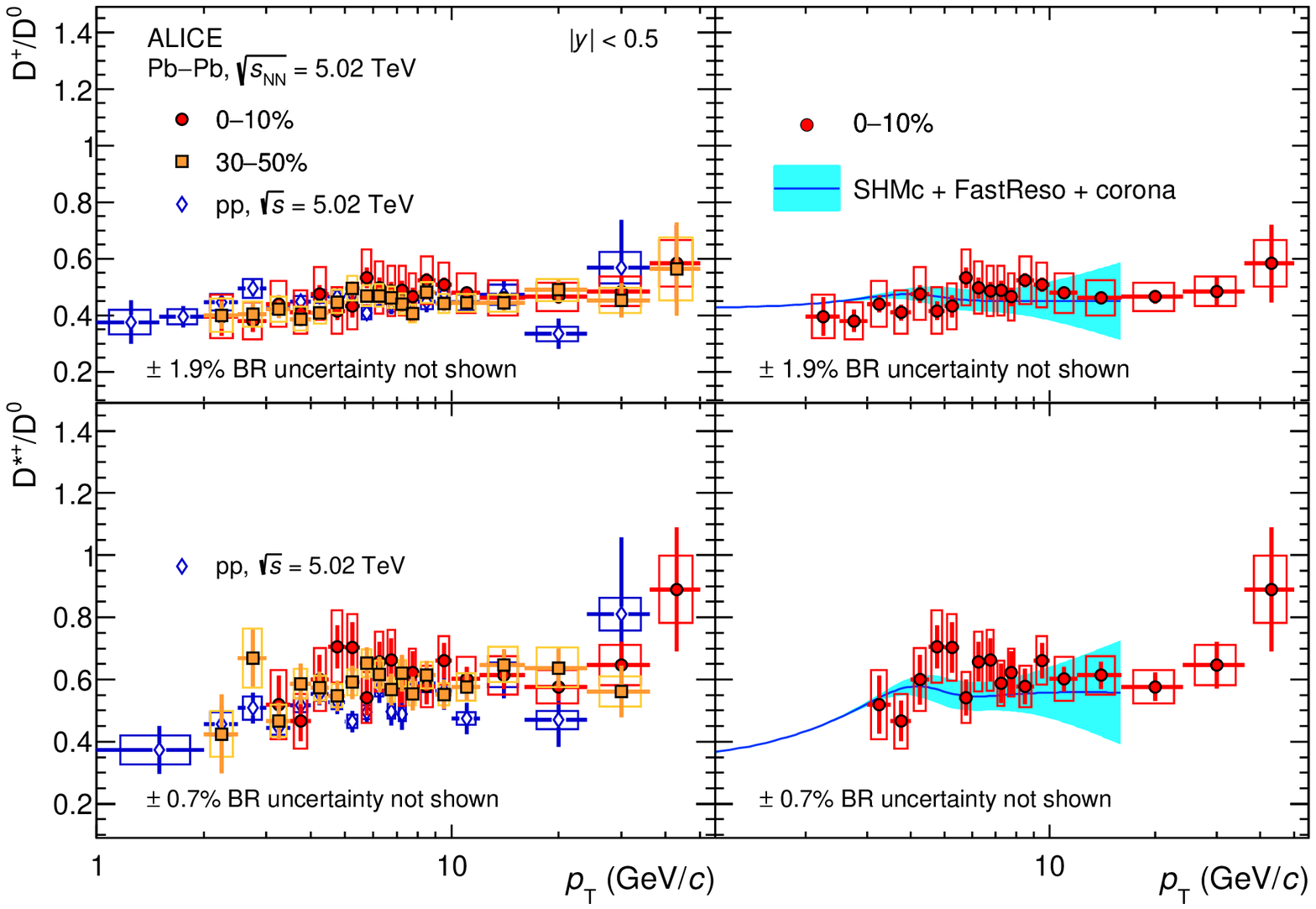}
\caption{$\Dplus/\Dzero$ (top left panel) and $\Dstar/\Dzero$ (bottom left panel) ratios as a function of $\pt$ in central and semicentral \PbPb collisions compared to the results obtained from pp collisions~\cite{Acharya:2019mgn}. The right panels show the ratios in central \PbPb collisions compared to the predictions from the statistical hadronisation model (SHMc)~\cite{Andronic:2019wva,Andronic:2021wva} with a core--corona approach and the FastReso package~\cite{Mazeliauskas:2018irt} for resonance decays. Statistical (bars) and systematic (boxes) uncertainties are shown.} 
\label{fig:Dmeson_ratios}
\end{center}
\end{figure}

Figure~\ref{fig:Dmeson_ratios} shows the $\pt$-dependent ratios of the production yields of prompt $\Dplus$ and $\Dstar$ mesons relative to the $\Dzero$ ones, in the left panels, compared with the values measured in pp collisions at $\sqrts = 5.02~\TeV$~\cite{Acharya:2019mgn}. The systematic uncertainties were propagated to the ratios, considering the contributions from the tracking efficiency and the beauty-hadron feed-down subtraction as fully correlated among the different D-meson species. The values obtained in \PbPb and pp collisions are compatible within uncertainties and this indicates there is no modification of the relative abundances of $\Dzero$, $\Dplus$, and $\Dstar$ mesons as a function of $\pt$ and for different centrality classes.
The $\Dplus/\Dzero$ and $\Dstar/\Dzero$ ratios are described within uncertainties by the GSI--Heidelberg statistical hadronisation model (SHMc)~\cite{Andronic:2019wva,Andronic:2021wva} (right panels of Fig.~\ref{fig:Dmeson_ratios}), which predicts the $\pt$ spectra of charm hadrons with a core--corona approach. The low-$\pt$ region is dominated by the core contribution described with a blast-wave function, while the corona contribution is more relevant at high $\pt$ and is parametrised from pp measurements. The rise predicted by the model, especially for $\Dstar/\Dzero$, are due to the different D-meson masses and to the collective radial expansion of the system.

The visible production yields of prompt $\Dzero$, $\Dplus$, and $\Dstar$ mesons in the two centrality classes were evaluated by integrating the $\dNdydpt$ over the $\pt$ intervals of the measurements. The results are reported in Table~\ref{tab:dNdy_visible}.
For the yield integration, the systematic uncertainty was propagated assuming all sources of uncertainty as fully correlated among $\pt$ intervals, except for the one on the raw-yield extraction, which was treated as uncorrelated because of the bin-to-bin variations of $S/B$ and background invariant-mass shape.

The production yield of prompt D mesons per unit of rapidity, $\dNdy$, was obtained by extrapolating (where needed) the visible cross section to the full $\pt$ range.
Since the $\pt$-differential yields of $\Dzero$ mesons in the 0--10\% and 30--50\% centrality classes were measured down to $\pt =$ 0, it was possible to obtain the $\dNdy$ at midrapidity for the first time in \PbPb collisions at the LHC without using models or assumptions to extrapolate the $\pt$-differential yield to unmeasured momentum range at low $\pt$.
Conversely, the $\dNdy$ of $\Dplus$ and $\Dstar$ mesons were obtained via an extrapolation procedure exploiting the measured $\pt$-differential ratios relative to the $\Dzero$ meson. 
The measured $\pt$-differential $\Dplus/\Dzero$ and $\Dstar/\Dzero$ ratios, shown in Fig.~\ref{fig:Dmeson_ratios}, were fit with two different functions (a constant and a constant plus logarithmic function) in order to extrapolate the ratios in the interval $0<\pt<2~\gevc$, where the yields could not be measured. The extrapolation in the high-$\pt$ range is negligible because the fraction of D-meson yield in central (semicentral) collisions with $\pt >$ 50 (36)$~\gevc$ is negligible.
The yields of $\Dplus$ and $\Dstar$ mesons in the interval $0<\pt<2~\gevc$ were obtained by multiplying the measured $\Dzero$ yield in this $\pt$ interval by the extrapolated ratio from the fit function. 
For the $\Dplus/\Dzero$ ratios, which show a flat trend as a function of $\pt$, the fit to a constant was used in the determination of the central value of the $\Dplus$ yield, while for the $\Dstar/\Dzero$ ratio the fit with a constant plus a logarithmic function was used. The systematic uncertainty due to the extrapolation was estimated by evaluating the yields using the other fit function and the shape of the SHMc predictions with normalisation parameter left free. The difference between the central value and the yields obtained with other functions for the extrapolation was assigned as systematic uncertainty and summed in quadrature to the uncertainty on the $\Dzero$ yield in the aforementioned $\pt$ interval.
The results for the prompt D-meson production yields per unit of rapidity $\dNdy$ in $|y|<$ 0.5 are reported in Table~\ref{tab:dNdy_010_integrated}. In the case of the $\Dzero$ meson, the result coincides with the visible yield because the measurement extends down to $\pt =$ 0 and the contribution of $\Dzero$ mesons with $\pt>$ 50 (36)$~\gevc$ is negligible.

\begin{table}[!t]
\renewcommand*{\arraystretch}{1.2}
\begin{center}
\caption{Visible production yield of prompt $\Dzero$, $\Dplus$, and $\Dstar$ mesons in $|y|<$ 0.5 in the 0--10\% and 30--50\% centrality classes of \PbPb collisions at $\sqrtsNN = 5.02~\TeV$.}
\begin{tabular}{c|c|c}
\toprule
 & Kinematic range & Visible production yield \\
\hline
\multicolumn{3}{c}{\rule{0pt}{2.6ex} 0--10\% centrality} \\
\hline
 \rule{0pt}{2.6ex} 
 $\Dzero$            & 0 $< \pt <$ 50 $\gevc$  & 6.819 $\pm$ 0.457 {\rm(stat.)} $^{+0.912}_{-0.936}$ {\rm(syst.)} $\pm$ 0.054 {\rm(BR)}\\
 \rule{0pt}{2.6ex} 
 $\Dplus$            & 2 $< \pt <$ 50 $\gevc$  & 0.992 $\pm$ 0.073 {\rm(stat.)} $^{+0.154}_{-0.155}$ {\rm(syst.)} $\pm$ 0.017 {\rm(BR)}\\
 \rule{0pt}{2.6ex}\rule[-1.2ex]{0pt}{0pt} 
 $\Dstar$            & 3 $< \pt <$ 50 $\gevc$  & 0.438 $\pm$ 0.037 {\rm(stat.)} $^{+0.084}_{-0.085}$ {\rm(syst.)} $\pm$ 0.005 {\rm(BR)} \\
\hline
\multicolumn{3}{c}{\rule{0pt}{2.6ex} 30--50\% centrality} \\
\hline
 \rule{0pt}{2.6ex} 
 $\Dzero$            & 0 $< \pt <$ 50 $\gevc$  & 1.275 $\pm$ 0.099 {\rm(stat.)} $^{+0.167}_{-0.173}$ {\rm(syst.)} $\pm$ 0.010 {\rm(BR)}\\
 \rule{0pt}{2.6ex} 
 $\Dplus$            & 2 $< \pt <$ 50 $\gevc$  & 0.179 $\pm$ 0.008 {\rm(stat.)} $^{+0.024}_{-0.024}$ {\rm(syst.)} $\pm$ 0.003 {\rm(BR)}\\
 \rule{0pt}{2.6ex}\rule[-1.2ex]{0pt}{0pt} 
 $\Dstar$            & 2 $< \pt <$ 36 $\gevc$  & 0.230 $\pm$ 0.023 {\rm(stat.)} $^{+0.038}_{-0.039}$ {\rm(syst.)} $\pm$ 0.002 {\rm(BR)}\\
\bottomrule
\end{tabular}
\label{tab:dNdy_visible}
\end{center}
\end{table}

\begin{table}[!tb]
\renewcommand*{\arraystretch}{1.2}
\begin{center}
\caption{Measured $\pt$-integrated yields of prompt $\Dzero$, $\Dplus$, and $\Dstar$ in $|y|<$ 0.5 in the 0--10\% and 30--50\% centrality classes of \PbPb collisions at $\sqrtsNN = 5.02~\TeV$. The right column reports the $\dNdy$ predicted by the GSI--Heidelberg statistical hadronisation model~\cite{Andronic:2019wva,Andronic:2021wva}.}
\begin{tabular}{c|l|c}
\toprule
 & \multicolumn{1}{|c|}{Measured $\dNdy$} & SHMc $\dNdy$ \\
\hline
\multicolumn{3}{c}{\rule{0pt}{2.6ex} 0--10\% centrality} \\
\hline
 \rule{0pt}{2.6ex} 
 $\Dzero$            & 6.819 $\pm$ 0.457 {\rm(stat.)} $^{+0.912}_{-0.936}$ {\rm(syst.)} $\pm$ 0.054 {\rm(BR)} & 6.42 $\pm$ 1.07 \\
 \rule{0pt}{2.6ex} 
 $\Dplus$            & 3.041 $\pm$ 0.073 {\rm(stat.)} $^{+0.154}_{-0.155}$ {\rm(syst.)} $\pm$ 0.052 {\rm(BR)} $^{+0.352}_{-0.618}$ {\rm(extrap.)} & 2.84 $\pm$ 0.47\\
 \rule{0pt}{2.6ex}\rule[-1.2ex]{0pt}{0pt} 
 $\Dstar$            & 3.803 $\pm$ 0.037 {\rm(stat.)} $^{+0.084}_{-0.085}$ {\rm(syst.)} $\pm$ 0.041 {\rm(BR)} $^{+0.854}_{-1.175}$ {\rm(extrap.)} & 2.52 $\pm$ 0.42\\
 \hline
\multicolumn{3}{c}{\rule{0pt}{2.6ex} 30--50\% centrality} \\
\hline
 \rule{0pt}{2.6ex} 
 $\Dzero$            & 1.275 $\pm$ 0.099 {\rm(stat.)} $^{+0.167}_{-0.173}$ {\rm(syst.)} $\pm$ 0.010 {\rm(BR)} & 1.06 $\pm$ 0.15 \\
 \rule{0pt}{2.6ex} 
 $\Dplus$            & 0.552 $\pm$ 0.008 {\rm(stat.)} $^{+0.024}_{-0.024}$ {\rm(syst.)} $\pm$ 0.009 {\rm(BR)} $^{+0.068}_{-0.114}$ {\rm(extrap.)} & 0.471 $\pm$ 0.069\\
 \rule{0pt}{2.6ex}\rule[-1.2ex]{0pt}{0pt} 
 $\Dstar$            & 0.663 $\pm$ 0.023 {\rm(stat.)} $^{+0.038}_{-0.039}$ {\rm(syst.)} $\pm$ 0.007 {\rm(BR)} $^{+0.149}_{-0.165}$ {\rm(extrap.)} & 0.419 $^{+0.065}_{-0.061}$\\
\bottomrule
\end{tabular}
\label{tab:dNdy_010_integrated}
\end{center}
\end{table}

In the right column the $\dNdy$ values predicted by the SHMc~\cite{Andronic:2019wva,Andronic:2021wva} are reported. In the SHMc approach, the yield of hadrons containing charm quarks can be calculated utilising as input values (i) the temperature $T_{\mathrm{chem}}$, (ii) the volume of the fireball at the chemical freeze-out estimated from a fit to light-flavour hadron yields, and (iii) the number of ${\rm c\overline{c}}$ pairs produced in the Pb--Pb collision. The latter was calculated from the charm-quark production cross section estimated from measurements in pp and \pPb\ collisions along with guidance from parameterisations of the nuclear modification of the PDFs~\cite{Andronic:2017pug}.
The SHMc calculations agree with the measurements within uncertainties, even though the data lie on the upper edge of the uncertainty band of the theoretical predictions, which is related to the uncertainty on the total charm production cross section used in the calculation.
It is useful to note that the charm production cross section in pp collisions, which was used to determine the charm content of the fireball in the SHMc calculations, is lower than the recent measurement reported in Ref.~\cite{ALICE:2021dhb} and therefore the yields from the SHMc would increase if the measured cross section were used as the input to the calculations.
However, a firmer conclusion on the SHMc predictions for charm hadrons will be drawn when the measured $\dNdy$ of $\Ds$, $\Lambda^{+}_{\rm{c}}$, and $\Jpsi$ will be available in Pb--Pb collisions and included in the comparison, since it is expected that the modifications of the hadronisation mechanisms in the presence of a QGP affect the relative abundances of different charm-hadron species~\cite{Kuznetsova:2006bh,Andronic:2003zv,Sorensen:2005sm,Lee:2007wr,Oh:2009zj,He:2012df,He:2019vgs}.

\subsection{Nuclear modification factor}
\begin{figure}[!t]
\begin{center}
\includegraphics[width=0.49\textwidth]{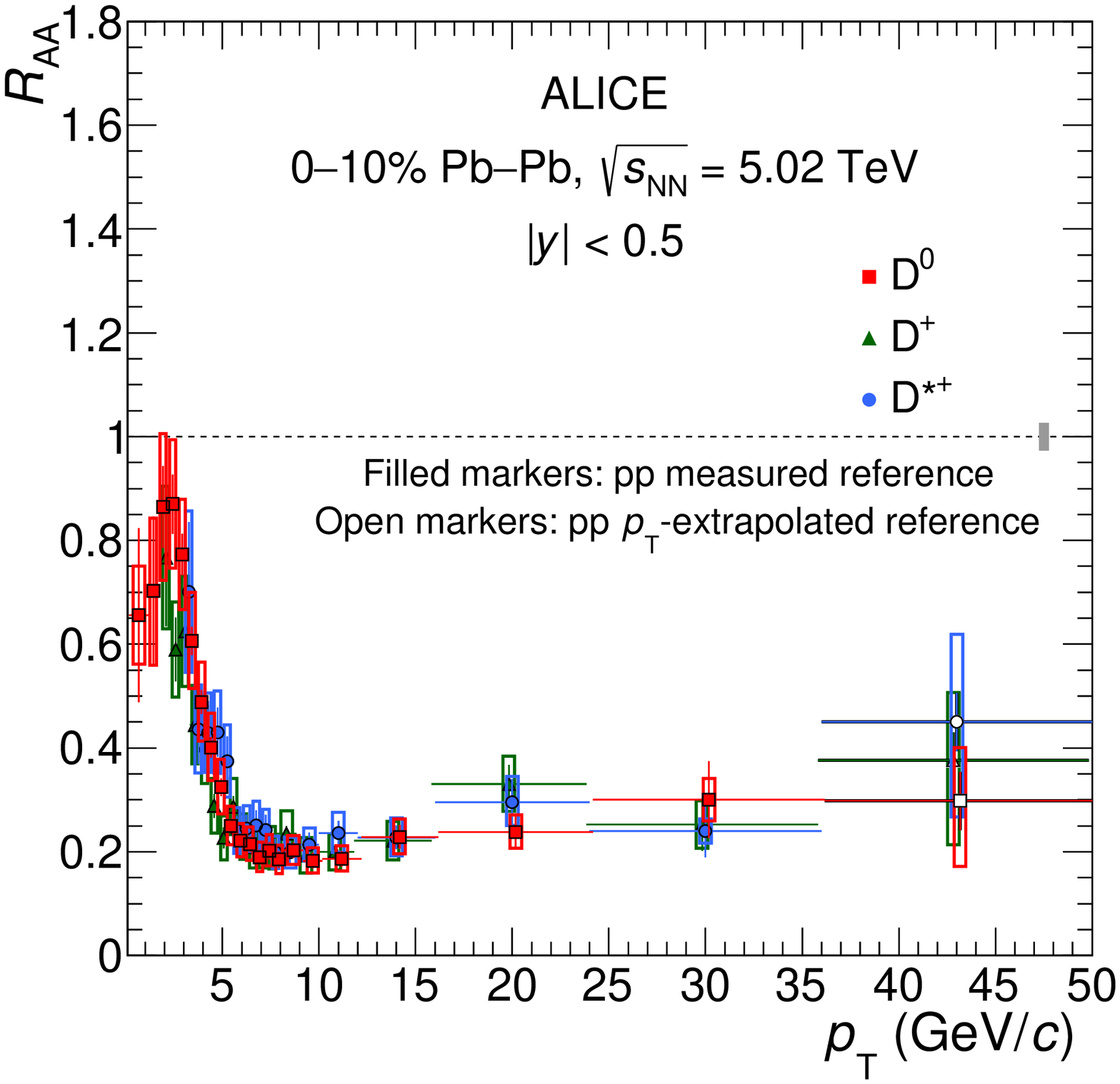}
\includegraphics[width=0.49\textwidth]{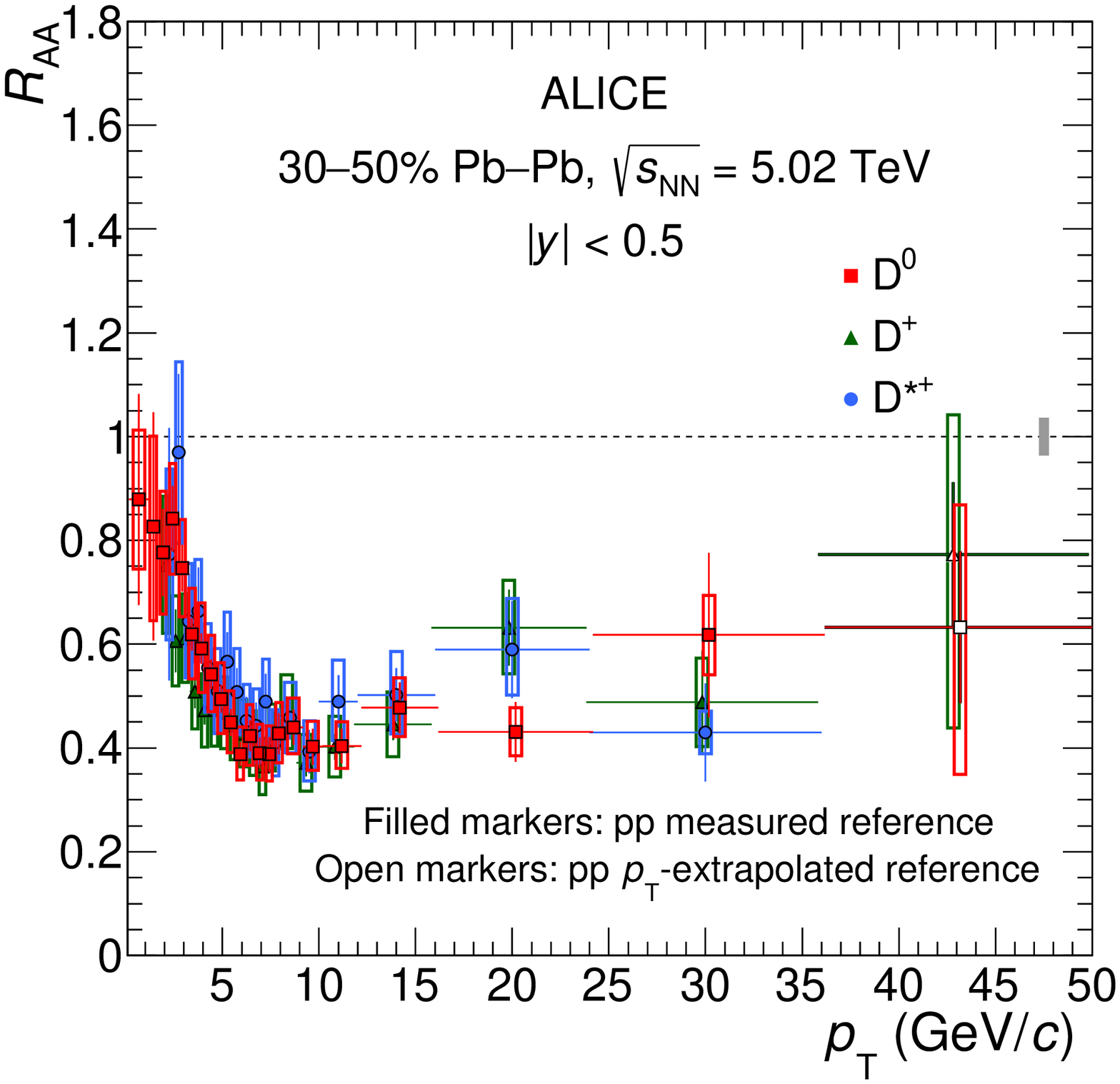}
\caption{$\RAA$ of prompt $\Dzero$, $\Dplus$, and $\Dstar$ mesons as a function of $\pt$ for the 0--10\% (left panel) and 30--50\% (right panel) centrality classes. Statistical (bars), systematic (boxes), and normalisation (shaded box around unity) uncertainties are shown. Filled markers are obtained with the measured pp reference~\cite{Acharya:2019mgn,Acharya:2021cqv}, empty markers with the $\pt$-extrapolated reference.} 
\label{fig:Dmeson_raa}
\end{center}
\end{figure}

The $\RAA$ of prompt $\Dzero$, $\Dplus$, and $\Dstar$ mesons was computed, as defined in Eq.~\ref{eq:Raa}, using the $\dNdydpt$ measured in \PbPb collisions and the pp reference at the same centre of mass energy reported in Fig.~\ref{fig:Dspectra}. The obtained results are shown in Fig.~\ref{fig:Dmeson_raa} for the two centrality classes. The nuclear modification factors of the three D-meson species are compatible among each other within statistical uncertainties.

The average $\RAA$ of $\Dzero$, $\Dplus$, and $\Dstar$ mesons was computed as a weighted average using the inverse of the quadratic sum of the relative statistical and uncorrelated systematic uncertainties as weights, in the $\pt$ intervals where more than one D-meson $\RAA$ value is available. 
The systematic uncertainties due to the raw-yield extraction and selection efficiency were considered as uncorrelated among different D-meson species and therefore they were used in the definition of the weights and propagated through the weighted average as uncorrelated, while the other sources of uncertainty (tracking efficiency, generated $\pt$ shape, and beauty hadron feed-down) were treated as fully correlated.

\begin{figure}[!t]
\begin{center}
\includegraphics[width=0.49\textwidth]{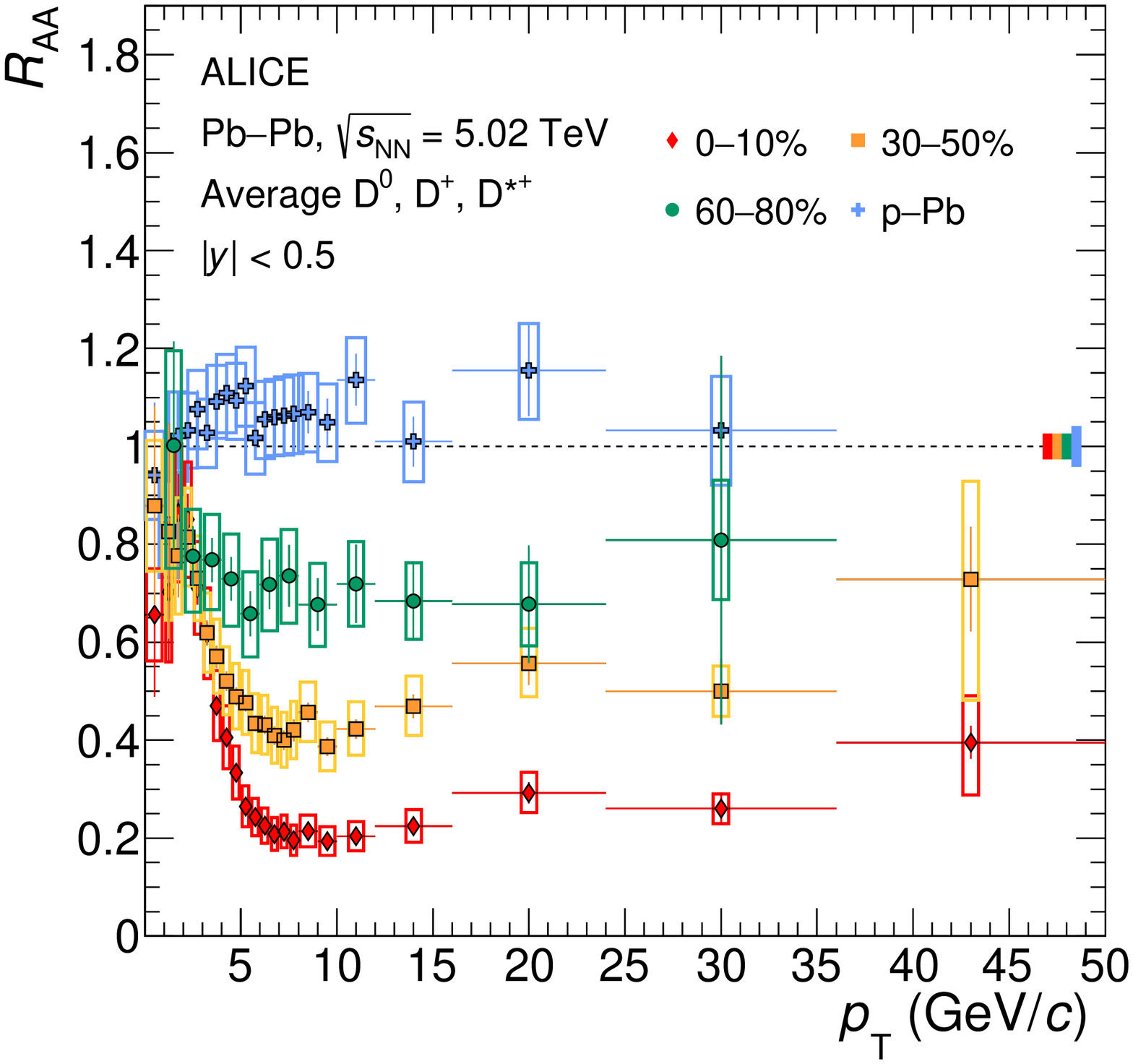}
\includegraphics[width=0.49\textwidth]{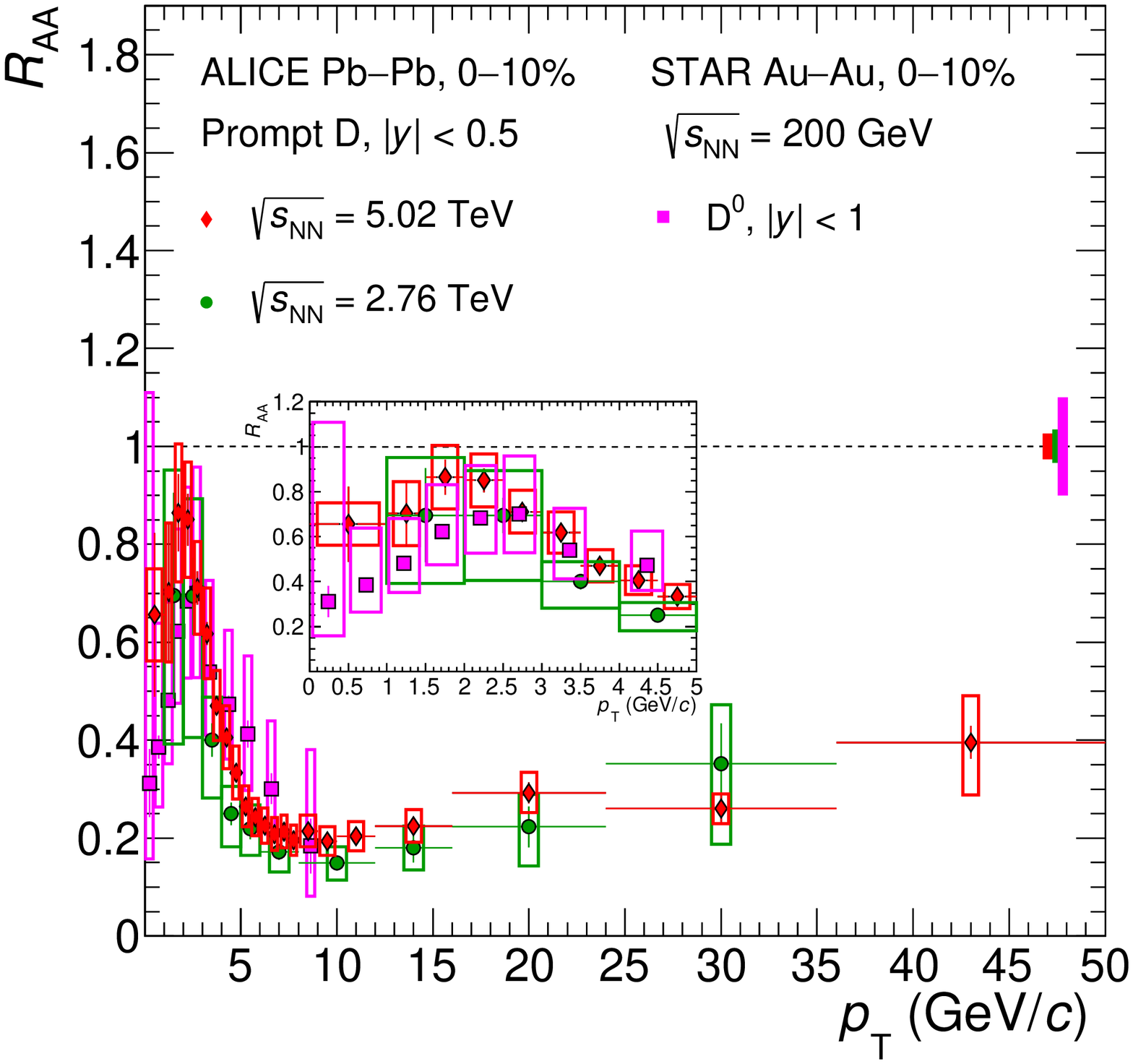}
\caption{Left panel: prompt D-meson $\RAA$ (average of $\Dzero$, $\Dplus$, and $\Dstar$) as a function of $\pt$ measured in \PbPb collisions at $\sqrtsNN = 5.02~\TeV$ (2018 data sample) in the 0--10\% and 30--50\% centrality classes compared with published results in the 60--80\% centrality class (2015 data sample)~\cite{Acharya:2018hre} and in \pPb collisions at the same centre-of-mass energy~\cite{Acharya:2019mno}.  Statistical (bars), systematic (boxes), and normalisation (shaded box around unity) uncertainties are shown. 
	Right panel: prompt D-meson $\RAA$ in the 10\% most central \PbPb collisions at $\sqrtsNN=~$5.02 TeV and 2.76 TeV~\cite{Adam:2015sza} compared to the $\Dzero$ $\RAA$ measured by the STAR collaboration in $\AuAu$ collisions at $\sqrtsNN=~$200 GeV~\cite{Adam:2018inb}.} 
\label{fig:Dmeson_avg_raa}
\end{center}
\end{figure}

The prompt D-meson average nuclear modification factors in the 0--10\% and 30--50\% centrality classes are shown in Fig.~\ref{fig:Dmeson_avg_raa} together with the $\RAA$ in the 60--80\% centrality class taken from Ref.~\cite{Acharya:2018hre}, which was measured using the sample of \PbPb collisions collected in 2015. The suppression increases from peripheral to central collisions. The $\RAA$ shows a minimum value at $\pt =$ 6--8 $\gevc$, corresponding to a suppression of the yields by a factor 5 and 2.5 with respect to the binary-scaled pp reference in the \mbox{0--10\%} and 30--50\% classes, respectively. The stronger suppression observed in central collisions is due to the increasing medium density, size, and lifetime of the fireball from peripheral to central collisions. Also shown in Fig.~\ref{fig:Dmeson_avg_raa} is the nuclear modification factor $\RpPb$ measured in $\pPb$ collisions at $\sqrtsNN = 5.02~\TeV$ taken from Ref.~\cite{Acharya:2019mno}, which is compatible with unity within uncertainties, confirming that the suppression observed in \PbPb collisions is to due final-state effects induced by the formation of a hot and dense QGP medium.

The right panel of Fig.~\ref{fig:Dmeson_avg_raa} shows the $\RAA$ of prompt D mesons measured by the ALICE collaboration in \PbPb collisions at $\sqrtsNN = 5.02~\TeV$ and 2.76 TeV~\cite{Adam:2015sza} and of $\Dzero$ mesons measured at RHIC by the STAR collaboration in \AuAu collisions at $\sqrtsNN=~$200 GeV~\cite{Adam:2018inb}. The results from ALICE at the two different energies are compatible within uncertainties. This similarity originates from the interplay between the higher medium temperature and density at 5.02 TeV, which causes a larger energy loss and a lower $\RAA$, and the harder $\pt$ distribution of charm quarks at 5.02 TeV, which would increase the $\RAA$ if the medium temperature were the same at the two collision energies~\cite{Djordjevic:2015hra}. The nuclear modification factor of $\Dzero$ mesons at RHIC energies shows a trend with $\pt$ similar to the one measured at the LHC, with a possible hint for a smaller $\RAA$ at low $\pt$ ($<2~\gevc$) and a larger $\RAA$ at high $\pt$ ($>4~\gevc$) in collisions at $\sqrtsNN=~$200 GeV compared to LHC energies. This difference could be due to dependence with collision energy of the charm-quark $\pt$ distributions, the initial-/final-state effects, and the medium properties. However, the rather large uncertainties prevent a firm conclusion on a $\sqrtsNN$ dependence of the $\RAA$ from being drawn.

\begin{figure}[!t]
\begin{center}
\includegraphics[width=0.65\textwidth]{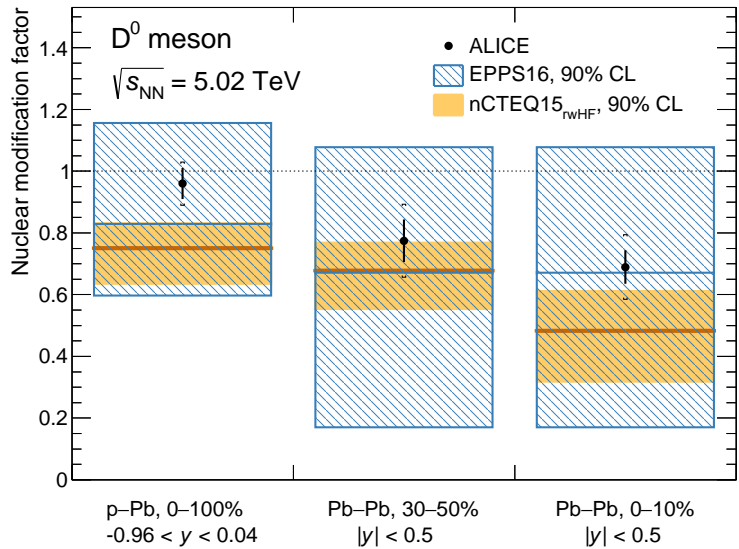}
\caption{$\pt$-integrated nuclear modification factors of prompt $\Dzero$ mesons measured in \pPb~\cite{Acharya:2019mno} and \PbPb collisions at $\sqrtsNN = 5.02~\TeV$. Statistical (bars) and systematic (brackets) uncertainties are shown. The results are compared with calculations at 90\% of confidence level of theoretical models nCTEQ15 (with Bayesian reweighting, see text for details)~\cite{Lansberg:2016deg,Kusina:2017gkz,Kusina:2020dki,Shao:2012iz,Shao:2015vga} and EPPS16~\cite{Eskola:2016oht,Helenius:2018uul} that include only initial-state effects.} 
\label{fig:D0_integrated_raa}
\end{center}
\end{figure}

In addition, the $\pt$-integrated $\RAA$ of prompt $\Dzero$ mesons at midrapidity was calculated from the $\pt$-integrated yield in \PbPb collisions reported above and the $\Dzero$ production cross section measured in pp collisions at $\sqrtsNN = 5.02~\TeV$~\cite{Acharya:2019mgn}. The results in the 0--10\% and 30--50\% centrality classes are:
\begin{equation}
\begin{split}
	R^{{\mathrm{prompt~D^0}}}_{{\mathrm{AA}}} (0\text{--}10\%) = 0.689 \pm 0.054~ ({\mathrm{stat.}}) ^{+0.104}_{-0.106}~({\mathrm{syst.}}), \\
	R^{{\mathrm{prompt~D^0}}}_{{\mathrm{AA}}} (30\text{--}50\%) = 0.775 \pm 0.069~({\mathrm{stat.}}) ^{+0.117}_{-0.120}~({\mathrm{syst.}}).	
\end{split}
\end{equation}
Figure~\ref{fig:D0_integrated_raa} shows the results obtained in \PbPb collisions compared with the nuclear modification factor $\RpPb$ measured in \pPb~collisions at the same centre-of-mass energy~\cite{Acharya:2019mno}. The total charm cross section is expected to scale with the number of binary collisions $\Ncoll$, as introduced in Section~\ref{sec:introduction}, and thus the $\RAA$ should be equal to one. However, the nuclear shadowing effect reduces the charm production in \PbPb (and \pPb) collisions with respect to pp interactions. In addition, the possible enhanced production of $\Ds$ and $\Lambda^{+}_{\mathrm{c}}$ due to the hadronisation via recombination is expected to further decrease the fraction of charm quarks that hadronise into $\Dzero$ mesons in \PbPb collisions compared to pp collisions~\cite{Kuznetsova:2006bh,Andronic:2003zv,Sorensen:2005sm,Lee:2007wr,Oh:2009zj,He:2012df,He:2019vgs}.
The measured $\pt$-integrated $\RAA$ is significantly below unity and this confirms the suppression of the $\Dzero$-meson yield in \PbPb collisions with respect to the binary-scaled pp reference due to shadowing and the possible modifications in the hadronisation mechanism. Conversely, the $\pt$-integrated $\Dzero$ $\RpPb$ is closer to unity, as expected from the smaller shadowing effects in \pPb~compared to \PbPb collisions (where it affects the nucleons of both the projectile and the target nuclei). The integrated $\RAA$ is also compared with perturbative QCD calculations of $\Dzero$-meson production including only initial-state effects modeled using two different sets of nuclear PDFs, namely nCTEQ15~\cite{Lansberg:2016deg,Kusina:2017gkz,Kusina:2020dki,Shao:2012iz,Shao:2015vga} and EPPS16~\cite{Eskola:2016oht,Helenius:2018uul}. The calculations with EPPS16 do not include the dependence of the shadowing on the impact parameter of the Pb--Pb collision and therefore they are the same in the central and semicentral event classes. The predictions with nCTEQ15 are obtained applying a Bayesian reweighting of the nuclear PDFs, which is constrained by measurements of heavy-flavour production in \pPb collisions at the LHC~\cite{Kusina:2017gkz}, and are labelled as nCTEQ15$_{\mathrm{rwHF}}$ in Fig.~\ref{fig:D0_integrated_raa}. They include a modelling of the centrality dependence of the nuclear modification of the PDFs. The uncertainty bands represent a 90\% confidence level uncertainty. In the nCTEQ15 case they are determined by considering three different factorisation scales in addition to the PDF uncertainties, with the scale variation constituting the main source of uncertainty, as described in Ref.~\cite{Kusina:2017gkz}. Both model calculations include only the initial-state effects due to the nuclear PDFs, but not the possible modifications of the relative abundances of different charm hadron species due to hadronisation via recombination. The measured $\RAA$ values lie on the upper edge of the theoretical predictions and this could be due to a smaller shadowing effect in the data with respect to the model expectations.

\subsection{Discussion: energy loss regime}
\label{sec:results_pQCD}

\begin{figure}[!t]
\begin{center}
\includegraphics[width=0.7\textwidth]{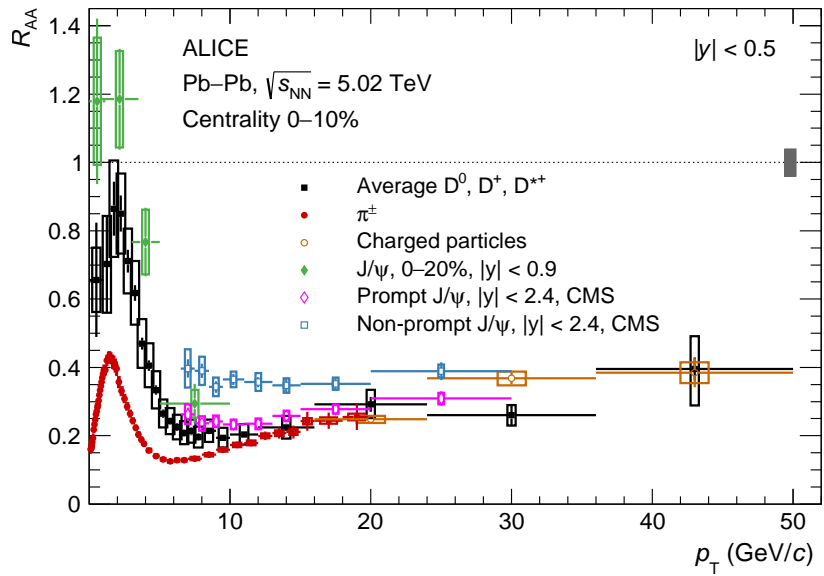}
\caption{Average $\RAA$ of prompt $\Dzero$, $\Dplus$, and $\Dstar$ mesons in the 0--10$\%$ centrality class compared to $\RAA$ of charged pions~\cite{Acharya:2019yoi}, charged particles~\cite{Acharya:2018qsh}, inclusive J/$\psi$ measured by ALICE~\cite{Acharya:2019lkh}, and of prompt and non-prompt J/$\psi$ from CMS~\cite{Sirunyan:2017isk} in \PbPb collisions at $\sqrtsNN = 5.02~\TeV$.}
\label{fig:D_pi_jpsi}
\end{center}
\end{figure}

The comparison of the nuclear modification factor of prompt D mesons with that of pions and of particles originating from beauty-hadron decays can provide essential insights into the characteristics of the in-medium parton energy loss, in particular on its predicted dependence on the colour charge and the quark mass.
To investigate possible differences with respect to light-flavour particles, the average $\RAA$ of prompt $\Dzero$, $\Dplus$, and $\Dstar$ mesons in the 0--10\% centrality class is compared in Fig.~\ref{fig:D_pi_jpsi} with that of charged pions~\cite{Acharya:2019yoi} and charged particles~\cite{Acharya:2018qsh}.
The charged-particle $\RAA$ is shown for $\pt>20~\gevc$ in order to extend the comparison to light-flavour particles up to the highest $\pt$ interval of the D-meson measurement.
The $\RAA$ of charged particles can be used in this comparison at high $\pt$ in place of the pion one because for $\pt>$ 8--10$~\gevc$. In this range, the nuclear modification factors of different light-flavour hadron species are found to be consistent with each other and the particle composition is compatible with that measured in pp collisions, that is dominated by pions~\cite{Abelev:2014laa,Acharya:2019yoi}.

The $\RAA$ of D mesons is larger than that of pions for $\pt<8~\gevc$, providing a clear evidence for a different nuclear modification factor of D mesons and light-flavour particles at low and intermediate $\pt$.
This difference originates from the interplay of several effects that concur in determining the magnitude and the $\pt$-dependence of the $\RAA$, so the intepretation in terms of different in-medium parton energy loss of charm quarks, light quarks, and gluons is not straighforward.
As pointed out in Ref.~\cite{Djordjevic:2013pba}, also in the presence of a colour-charge and quark-mass dependent energy loss, similar values of D-meson and pion $\RAA$ are expected at high $\pt$ ($\gtrsim$ 8$~\GeV/c$) due to the harder $\pt$ distribution and the harder fragmentation function of charm quarks compared to those of light quarks and gluons.
At low $\pt$, where a large difference is observed between the D-meson and pion $\RAA$'s, it should be considered that the pion yield can have a significant contribution from soft production up to transverse momenta of about 3--4$~\gevc$ due to the strong radial flow at LHC energies. This soft component, which is not present in the D-meson production, does not scale with the number of binary nucleon--nucleon collisions.
In addition, effects due to the radial flow and the hadronisation can affect D-meson and light-hadron yields differently at a given $\pt$, playing a role in the interpretation of their different $\RAA$ at low and intermediate $\pt$.
For instance, an enhanced production of $\Ds$ mesons and charm baryons in \PbPb  collisions due to recombination would imply a reduction of the fraction of charm quarks hadronising into non-strange meson species compared to pp collisions.

\begin{figure}[!t]
\begin{center}
\includegraphics[width=\textwidth]{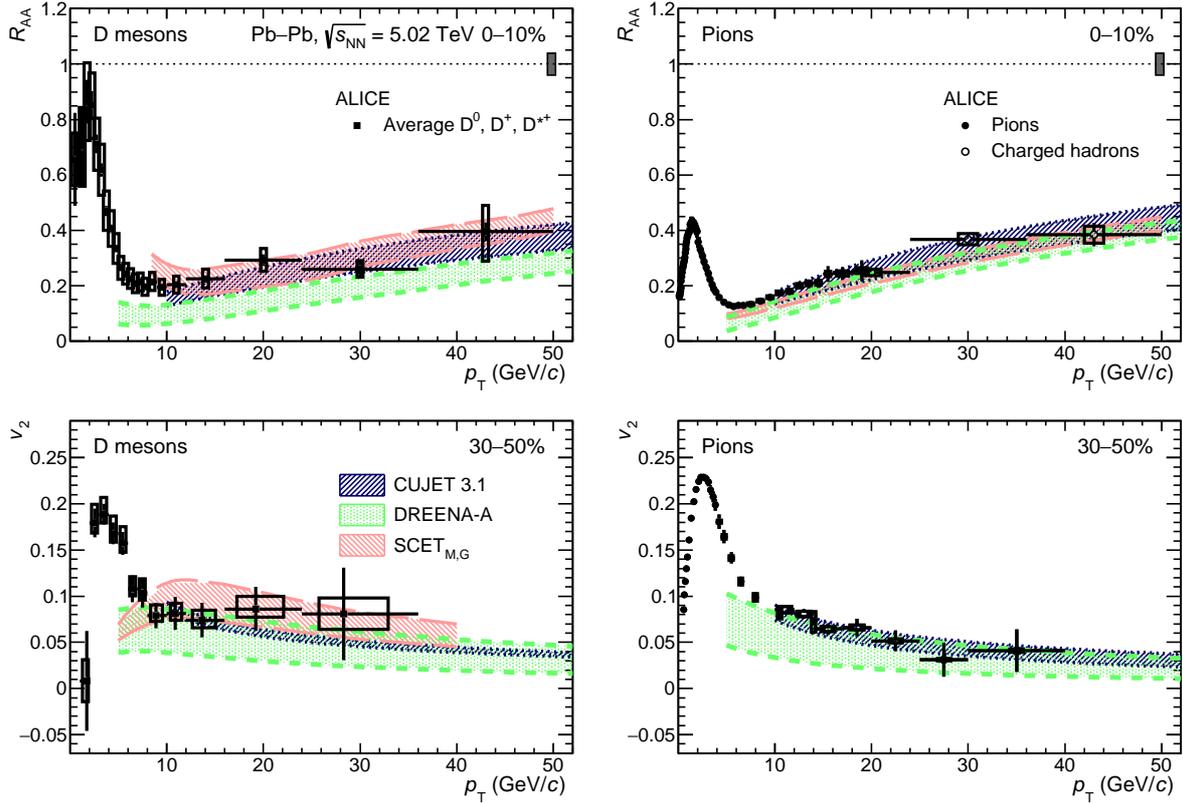}
\caption{Average $\RAA$ (top panels) and $v_2$ (bottom panels) of prompt $\Dzero$, $\Dplus$, and $\Dstar$ mesons~\cite{Acharya:2020pnh} and pions (charged hadrons)~\cite{Acharya:2019yoi,Acharya:2018qsh,Acharya:2018zuq}  compared with predictions from models based on pQCD calculations, namely CUJET3.1~\cite{Xu:2014tda,Xu:2015bbz,Shi:2019nyp}, DREENA-A~\cite{Stojku:2020tuk}, and ${\rm SCET_{\rm M,G}}$~\cite{Kang:2016ofv}.}
\label{fig:D_pi_models}
\end{center}
\end{figure}

A quantitative understanding of the parton in-medium energy-loss from the $\RAA$ measurements needs therefore to be carried out via comparisons with model calculations. In particular, in the high-$\pt$ region, where effects due to radial flow and modifications in the hadronisation mechanisms are expected to be negligible, models based on perturbative QCD (pQCD) calculations of high-$\pt$ parton energy loss are expected to describe the data.
The $\RAA$ and the elliptic flow $v_2$ of prompt D mesons (taken from Ref.~\cite{Acharya:2020pnh}) and pions (taken from Ref.~\cite{Acharya:2019yoi, Acharya:2018zuq}) in the 0--10\% and 30-50\% centrality classes are compared in Fig.~\ref{fig:D_pi_models} to three of these models, namely CUJET3.1~\cite{Xu:2014tda,Xu:2015bbz,Shi:2019nyp}, DREENA-A~\cite{Stojku:2020tuk,Zigic:2018ovr,Zigic:2018smz}, and ${\rm SCET_{M,G}}$~\cite{Kang:2016ofv}.
The DREENA-A is a numerical framework that calculates the medium-modified distribution of high-$\pt$ hadrons based on a dynamical energy-loss formalism coupled with a modelling of initial parton momentum distributions, a full 3+1D hydrodynamic evolution of the medium, and fragmentation functions. 
The dynamical energy-loss formalism~\cite{Djordjevic:2009cr,Djordjevic:2008iz,Djordjevic:2006tw} is a model of jet-medium interactions incorporating collisional and radiative energy loss mechanisms in a QCD medium of finite size and temperature composed of dynamical scattering centers. It includes a colour-charge and quark-mass dependence as well as finite magnetic mass effects and running strong coupling constant.
The CUJET3.1 framework provides a calculation of jet energy loss in a QCD medium described with relativistic viscous hydrodynamics.
The jet-medium interactions are based on the DGLV opacity expansion model~\cite{Gyulassy:2000fs,Gyulassy:2000er, Djordjevic:2003zk} including both inelastic and elastic scatterings with their colour-charge and quark-mass dependence, and taking into account interactions with both chromo-electric and magnetic charges of the medium.
The ${\rm SCET_{\rm M,G}}$ model implements medium-induced gluon radiation via modified splitting functions. It is based on a soft-collinear effective theory (SCET~\cite{Bauer:2000yr,Bauer:2001yt}) describing the parton shower formed in the vacuum via soft and collinear splittings.
This effective theory was extended to the case of massless~\cite{Idilbi:2008vm,Ovanesyan:2011xy} and massive~\cite{Kang:2016ofv} quarks propagating in strongly-interacting matter by including additional splitting processes induced by the interactions of the incident parton with the QCD medium mediated by Glauber gluon exchange.
These medium-induced splittings are calculated to first order in the opacity series expansion, thus limiting the applicability of these calculations to the high $\pt$ region.

The considered models provide a fair description of both the $\RAA$ and the $v_2$ of D mesons and pions for $\pt > 10~\gevc$, where radiative energy loss is expected to be the dominant interaction mechanism.
This suggests that the dependences of radiative energy loss on the colour charge and the quark mass of the hard-scattered parton, as well as on the path length in the hot and dense medium are reasonably well described in these calculations.

The comparison of the nuclear modification factors of particles originating from charm and beauty quarks is also reported in Fig.~\ref{fig:D_pi_jpsi}, to provide insight into the predicted quark-mass dependence of parton energy loss. In particular, the $\RAA$ of inclusive $\Jpsi$ mesons measured by ALICE~\cite{Acharya:2019lkh} at midrapidity in the 0--20\% centrality class and those of prompt and non-prompt $\Jpsi$ from the CMS collaboration~\cite{Sirunyan:2017isk} are shown. 

The $\RAA$ of prompt D mesons in the 0--10\% centrality class is lower than that of non-prompt $\Jpsi$ mesons from beauty hadron decays measured in the same centrality interval.
This difference provides an indication for the predicted quark-mass dependence of in-medium energy loss, even though a proper interpretation of the different nuclear modification factors in terms of energy loss requires a full modelling of the initial momentum distributions of charm and beauty quarks, of the heavy-quark fragmentation, and of the beauty-hadron decay kinematics.
The model from Ref.~\cite{Djordjevic:2016vfo}, which includes all these effects, provides a good description of the measurements of prompt D and non-prompt $\Jpsi$ mesons.
In this model, the large difference in the $\RAA$'s in the $\pt$ interval around 10~$\gevc$ is mostly due to the different in-medium energy loss of c and b quarks, as the effects of initial distributions, quark fragmentation, and beauty-hadron decays are found to be small.

The comparison of the nuclear modification factors of prompt D mesons and prompt (inclusive) $\Jpsi$ mesons, also shown in Fig.~\ref{fig:D_pi_jpsi}, provides insight into the interplay of different QGP medium effects in the charm sector, where they are expected to affect differently the production of open charm and charmonium states.
At high momentum ($\pt \gtrsim 10~\gevc$), the $\RAA$ of prompt D mesons is compatible within uncertainties with that of prompt $\Jpsi$ mesons, as well as with that of light-flavour hadrons. This may suggest that in this $\pt$ region the yield of charmonia has a significant contribution from production within the parton shower originating from the splitting of a hard-scattered gluon~\cite{Aaij:2017fak,Bain:2017wvk}, which experiences in-medium energy loss before fragmenting~\cite{Spousta:2016agr,Arleo:2017ntr}.
At lower $\pt$ ($2 \lesssim \pt \lesssim 4~\gevc$), the results for inclusive $\Jpsi$, which are dominated by the prompt contribution, show a magnitude and a trend of the $\RAA$ similar to that of D mesons.
In this region, $\Jpsi$ production is likely dominated by recombination of c and $\overline{\rm c}$ quarks in the medium either at the phase boundary or during the QGP expansion~\cite{Adam:2015isa,Acharya:2019lkh}, similarly to the recombination of charm and light quarks forming D mesons.
The similar $\RAA$ of $\Jpsi$ and D mesons in this momentum region may signal the dominant contribution of hadronisation via recombination after the interactions of the charm quarks with the QGP medium constituents~\cite{Zhou:2014kka,Du:2015wha,He:2011qa,Cao:2019iqs}, with possible small differences arising from the different kinematics involved in the charm-quark coalescence with $\overline{\rm c}$ and light antiquarks.

\subsection{Discussion: transport models and an estimate of the spatial diffusion coefficient}
\label{sec:results_transport}
\begin{figure}[!t]
\begin{center}
\includegraphics[width=1.05\textwidth]{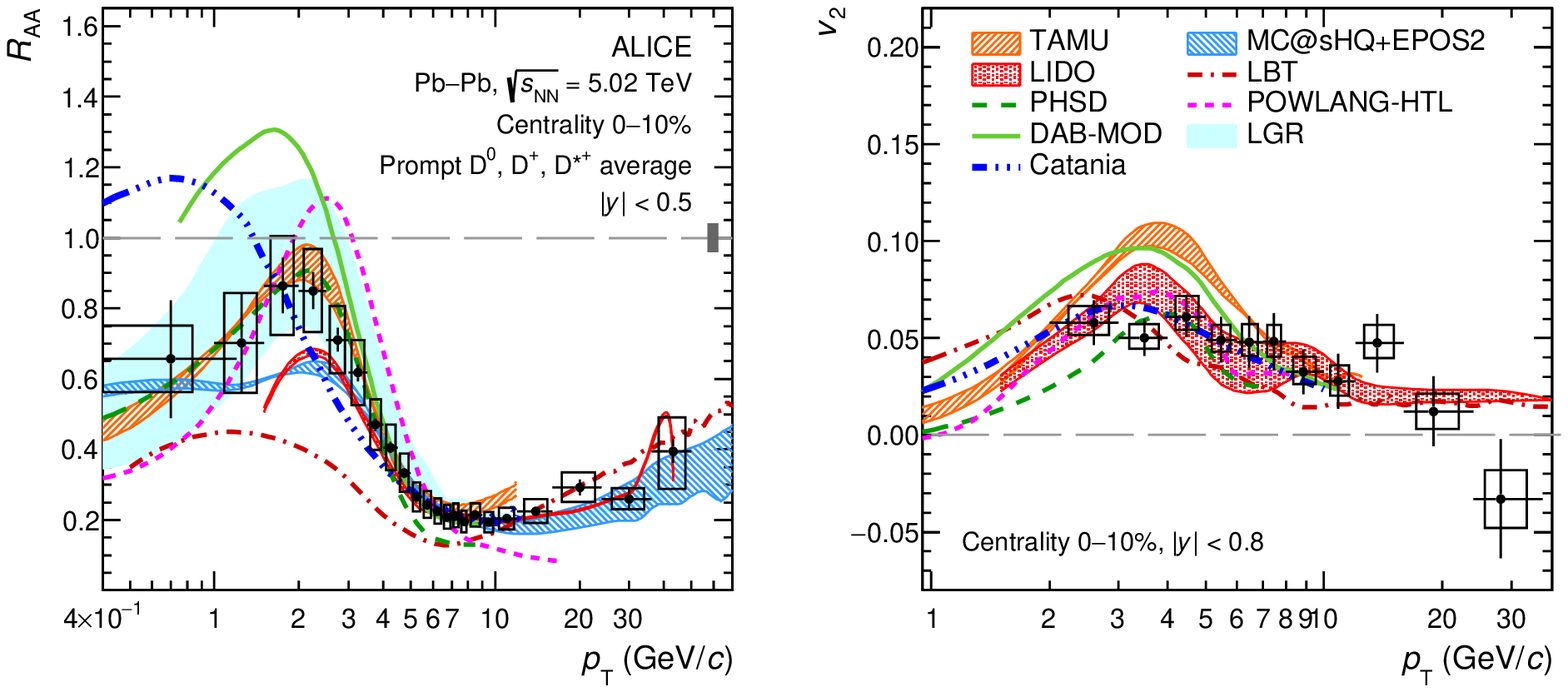}
\includegraphics[width=1.05\textwidth]{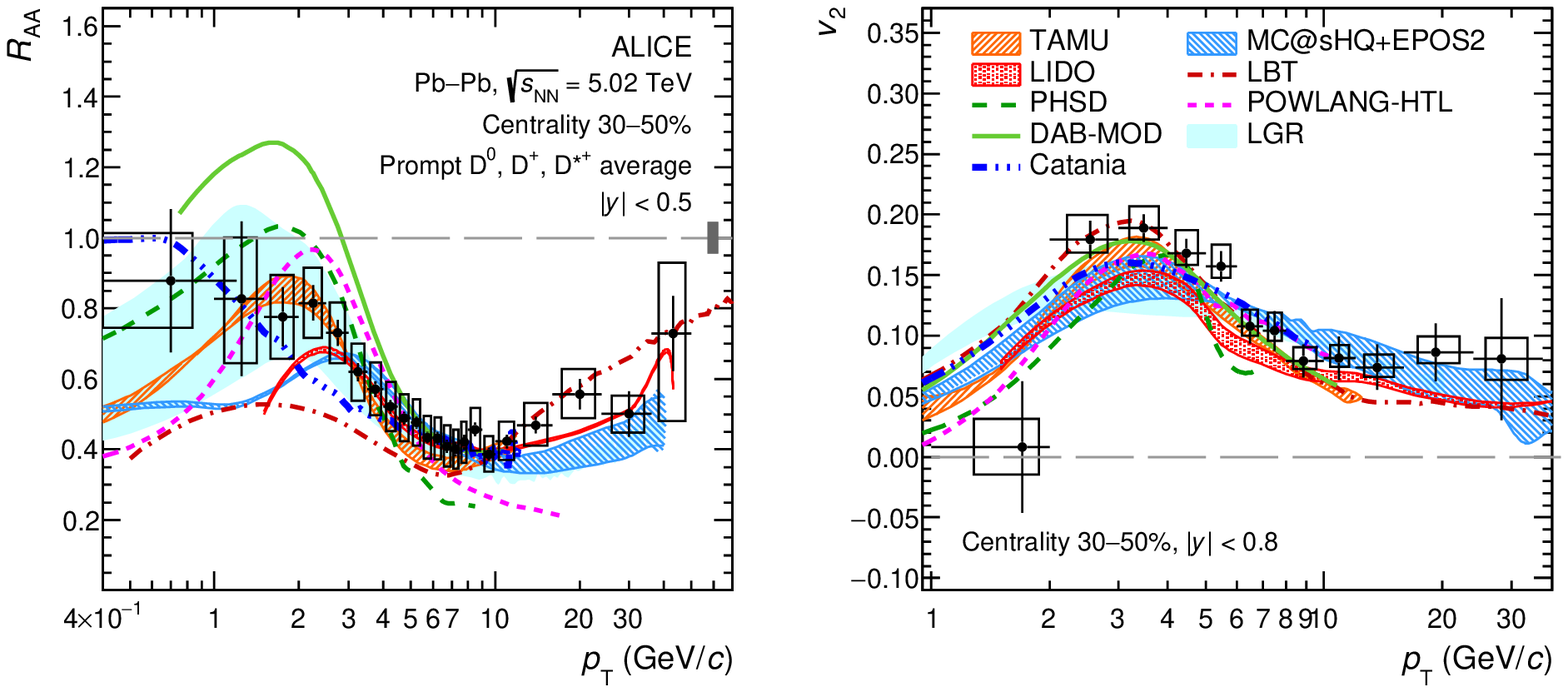}

\caption{Average $\RAA$ (left) and elliptic flow $v_{2}$~\cite{Acharya:2020pnh} (right) of prompt $\Dzero$, $\Dplus$, and $\Dstar$ mesons in the 0--10$\%$ (top) and 30--50$\%$ (bottom) centrality classes compared with predictions of models implementing the charm-quark transport in a hydrodynamically expanding medium~\cite{Nahrgang:2013xaa,Katz:2019fkc,Beraudo:2014boa,Beraudo:2017gxw,Cao:2016gvr,Cao:2017hhk,He:2019vgs,Li:2019lex,Scardina:2017ipo,Plumari:2019hzp,Ke:2018jem,Song:2015sfa}.} 
\label{fig:Dmeson_transport}
\end{center}
\end{figure}

As discussed in Section~\ref{sec:introduction}, the charm-hadron yields and angular distributions at low and intermediate $\pt$ are sensitive to the diffusion and the possible thermalisation of charm quarks in the medium. The comparison of the measured D-meson $\RAA$ and $v_2$ to models implementing charm-quark transport in a hydrodynamically expanding QGP could therefore provide insight into the interactions of heavy quarks with the medium, constraining in particular the spatial diffusion coefficient, which is the relevant transport coefficient in the diffusion regime. In the left panels of Fig.~\ref{fig:Dmeson_transport} the measured nuclear modification factor of prompt D mesons for the 0--10$\%$ (top) and 30--50$\%$ (bottom) centrality class, respectively, is compared with various predictions from models implementing charm-quark transport in a hydrodynamically expanding medium~\cite{Nahrgang:2013xaa,Katz:2019fkc,Beraudo:2014boa,Beraudo:2017gxw,Cao:2016gvr,Cao:2017hhk,He:2019vgs,Li:2019lex,Scardina:2017ipo,Plumari:2019hzp,Ke:2018jem,Song:2015sfa}. In particular, the TAMU~\cite{He:2019vgs}, POWLANG-HTL~\cite{Beraudo:2014boa,Beraudo:2017gxw}, PHSD~\cite{Song:2015sfa}, and Catania~\cite{Scardina:2017ipo,Plumari:2019hzp} models describe the interactions of the charm quarks with the medium constituents solely via collisional processes, while the MC@sHQ+EPOS2~\cite{Nahrgang:2013xaa}, DAB-MOD~\cite{Katz:2019fkc}, LBT~\cite{Cao:2016gvr,Cao:2017hhk}, LGR~\cite{Li:2019lex}, and LIDO~\cite{Ke:2018jem} calculations include also radiative processes. All the models, except for DAB-MOD, include initial-state effects by using nuclear PDFs (nPDFs) in the calculation of the initial $\pt$ distributions of charm quarks. A contribution of hadronisation via quark recombination, in addition to charm-quark fragmentation, is included in all theoretical predictions. The theoretical uncertainties, where available, are displayed with a coloured band.

Most of the models capture the measured $\pt$ trend and magnitude of the nuclear modification factor and provide a good description of the data in the $\pt$ interval 6 $< \pt <$ 10$~\gevc$ where the minimum of the $\RAA$ is observed, but many of them show significant deviations from the measurements at low $\pt$ ($\pt\lesssim$ 4--5$~\gevc$). In particular, the LBT~\cite{Cao:2016gvr,Cao:2017hhk}, LIDO~\cite{Ke:2018jem}, MC@sHQ+EPOS2~\cite{Nahrgang:2013xaa}, and Catania~\cite{Scardina:2017ipo,Plumari:2019hzp} calculations tend to underestimate the results for 2 $< \pt <$ 4$~\gevc$ in both centrality classes, while the PHSD~\cite{Song:2015sfa} prediction is above the measured $\RAA$ in semicentral collisions for the same $\pt$ range. The POWLANG-HTL~\cite{Beraudo:2014boa,Beraudo:2017gxw} calculation, instead, predicts a narrower and more pronounced radial flow peak at low $\pt$ as compared to the measured one for the 10\% most central \PbPb collisions. Most of the models underestimate the measured $\RAA$ in both centrality classes for $\pt <$ 1.5$~\gevc$, with the exception of Catania~\cite{Scardina:2017ipo,Plumari:2019hzp} calculations, which tend to overpredict the measured $\RAA$ at low $\pt$ in central and semicentral collisions and PHSD predictions, which overestimate the $\RAA$ in the 30--50\% centrality class. Nevertheless, it is important to remark that the $\RAA$ at low $\pt$ is sensitive not only to the modelling of the charm-quark interactions with the medium but also to the parametrisation of the nPDFs used in the calculations and to the hydrodynamical description of the underlying medium.

More stringent constraints to the implementation of the heavy-quark interactions with the medium constituents can be provided by the simultaneous comparison of $\RAA$ and $v_2$ measurements in \PbPb collisions at $\sqrtsNN = 5.02~\TeV$~\cite{Acharya:2020pnh} with models, as reported in Fig.~\ref{fig:Dmeson_transport}. All the models describe reasonably well the $v_2$ in the most central collisions, while they tend to underestimate the measured points in the 2 $< \pt <$ 6$~\gevc$ interval for the 30--50\% centrality class, except for LBT which reproduces well the measured $v_2$ but misses completely the $\RAA$ in the same $\pt$ range.  
In this $\pt$ region, the measured $v_2$ originates predominantly from the charm-quark interactions with the QGP constituents, which impart the flow of the medium to the heavy quarks, and from the hadronisation via recombination, which enhances the charm-hadron $v_2$ with respect to the one of the charm quark because the D meson picks up the $v_2$ of the light quark. 
Similarly, the peak observed in the $\RAA$ for 2 $< \pt <$ 6$~\gevc$ is also due to the interplay between the diffusion and recombination of the charm quarks with the medium constituents. Thus, the measurements of the $\RAA$ and $v_2$ in this $\pt$ region are particularly sensitive to quark diffusion and thermalisation with the medium, and to the hadronisation mechanisms.

The simultaneous description of $\RAA$ and $v_2$ is challenging for the models and therefore the data have the potential to constrain the model ingredients and parameters.
The global agreement between the measured $\RAA$ and the theoretical models was evaluated by computing a $\chi^{2}/$ndf, as done in Refs.~\cite{Acharya:2017qps, Acharya:2020pnh}. The statistical and systematic uncertainties (treating separately the contributions correlated and uncorrelated among $\pt$ intervals) were considered in the calculation together with the theoretical ones, when available. Since the upper $\pt$ limit of the predictions is different for each model, the $\chi^{2}/$ndf was computed in the 0 $< \pt <$ 8$~\gevc$ interval, which is common among all the models except for the LIDO predictions which start from 1.5$~\gevc$. Therefore, the $\chi^{2}/$ndf computation for LIDO was performed excluding the first two $\pt$ intervals of the $\RAA$. This low-$\pt$ range provides high sensitivity to charm-quark diffusion and hadronisation in the QGP. The $\chi^{2}/$ndf values are reported in Table~\ref{tab:chi2}. 
The large spread in the computed $\chi^{2}/$ndf is not only due to the improved precision of the measurement, but also to the differences among the theoretical models. They do not only differ in terms of the interaction of charm quarks with the medium, as previously highlighted, but also in terms of the considered nuclear PDFs, the bulk evolution of the medium (i.e. ideal or viscous hydrodynamics), the charm hadronisation mechanism (i.e. fragmentation and/or recombination), and whether a hadronic phase is included or not. 
In particular, the charm-quark recombination mechanism is implemented with different approaches in the various models: most of the models use an instantaneous recombination based on the Wigner function formalism~\cite{Dover:1991zn}, while TAMU implements a resonance-recombination model~\cite{Ravagli:2007xx}, PHSD a dynamical coalescence via a Monte Carlo approach~\cite{Song:2015sfa}, and POWLANG an in-medium string formation approach~\cite{Beraudo:2014boa}.
The $\RAA$ predictions are deeply influenced by these additional model ingredients, other than the transport properties of the medium.
Therefore, a rather mild requirement on the data-to-model consistency, namely $\chi^{2}/$ndf~$<$~5, was applied to select the models considered for the estimation of the heavy-quark spatial diffusion coefficient. With this criterion, the selected models are TAMU~\cite{He:2019vgs}, MC@sHQ+EPOS2~\cite{Nahrgang:2013xaa}, LIDO~\cite{Ke:2018jem}, LGR~\cite{Li:2019lex}, and  Catania~\cite{Scardina:2017ipo,Plumari:2019hzp}. 
A similar analysis was performed for the elliptic and triangular flow~\cite{Acharya:2020pnh}. 
In order to further constrain the description of heavy-quark transport in the medium and the spatial diffusion coefficient $D_s$, the results of the $\chi^{2}/$ndf analysis for the $\RAA$ model calculations were combined with those obtained for the elliptic and triangular flow in Ref.~\cite{Acharya:2020pnh}. Note that some theoretical predictions for the $v_2$ were updated after the publication of Ref.~\cite{Acharya:2020pnh}, namely LBT~\cite{Cao:2016gvr,Cao:2017hhk}, LIDO~\cite{Ke:2018jem}, LGR~\cite{Li:2019lex}, PHSD~\cite{Song:2015sfa}, and TAMU~\cite{He:2019vgs}. Therefore the $\chi^{2}/$ndf of these predictions with respect to the measured $v_2$ and $v_3$ was re-computed. 
The outcome did not change significantly with respect to Ref.~\cite{Acharya:2020pnh}, with the only exception of LIDO~\cite{Ke:2018jem} for which the updated predictions provide a better description of the data with respect to the old ones, resulting in a $\chi^{2}/$ndf~$<$~2. Thus, the LIDO model is included among those providing a good description of $v_2$ and $v_3$, while it was not considered in Ref.~\cite{Acharya:2020pnh}.  
The models that describe reasonably well both $\RAA$ (with $\chi^{2}/$ndf~$<$~5) and $v_2$ and $v_3$ (with $\chi^{2}/$ndf~$<$~2) are TAMU~\cite{He:2019vgs}, MC@sHQ+EPOS2~\cite{Nahrgang:2013xaa}	, LIDO~\cite{Ke:2018jem}, LGR~\cite{Li:2019lex}, and Catania~\cite{Scardina:2017ipo,Plumari:2019hzp}. These models use a value of heavy-quark spatial diffusion coefficient in the range $1.5 < 2\pi D_\mathrm{s} T_\mathrm{c} < 4.5$ at the pseudocritical temperature $T_\mathrm{pc} = 155~\MeV$~\cite{Borsanyi:2010bp}. 
According to lQCD calculations the $2\pi D_\mathrm{s} T_\mathrm{c}$ lies between 2 and 6~\cite{Kaczmarek:2014jga,Ding:2012sp,Banerjee:2011ra}, and these results are also in agreement with the $D_s$ range estimated from the $v_2$ measurements by ALICE (1.5--7 from Ref.~\cite{Acharya:2020pnh}), STAR (2--12 from Ref.~\cite{Adamczyk:2017xur}), and PHENIX (value of $\sim$3 from Ref.~\cite{Adare:2010de}).
The inclusion of the $\RAA$ in the $\chi^{2}/$ndf improved the constraint on the spatial diffusion coefficient with respect to the range reported in Ref.~\cite{Acharya:2020pnh}.
It is however important to remark that this coefficient is not the only key parameter of the models describing the heavy-quark transport in an expanding medium, but there are other ingredients (such as the parameters of the underlying hydrodynamics, the modelling of the hadronisation, the description of the interactions in the hadronic phase, amongst others, see e.g. Refs.~\cite{Rapp:2018qla,Cao:2018ews}) playing an important role.

\begin{table}[!tb]
\caption{Summary of the $\chi^2/\mathrm{ndf}$ values obtained in $0<\pt<8~\gevc$ for the different model predictions compared with the measured D-meson $\RAA$.}
\centering
\renewcommand*{\arraystretch}{1.2}
\begin{tabular}[t]{l|c c c}
\toprule
\multirow{3}{*}{Model} & \multicolumn{3}{c}{$\chi^2/\mathrm{ndf}$}\\

\cline{2-4}
& \multicolumn{2}{c}{$\RAA$} & \multirow{2}{*}{global} \\
\cline{2-3}
&  0--10\% & 30--50\% & \\
\hline
\hline
Catania~\cite{Scardina:2017ipo,Plumari:2019hzp}		& 94.9$/$15 		& 48.9$/$15 	& 143.8$/$30 \\
DAB-MOD~\cite{Katz:2019fkc}							& 110.2$/$15 	& 123.9$/$15  	& 234.1$/$30 \\
LBT~\cite{Cao:2016gvr,Cao:2017hhk}					& 342.6$/$15 	& 69.2$/$15  	& 411.8$/$30 \\
LIDO~\cite{Ke:2018jem}								& 31.8$/$13		& 14.5$/$13  	& 46.4$/$26 \\
LGR~\cite{Li:2019lex}								& 4.7$/$15		& 4.5$/$15   	& 9.2$/$30  \\
MC@sHQ+EPOS2~\cite{Nahrgang:2013xaa}					& 31.8$/$15		& 24.8$/$15  	& 56.6$/$30  \\
PHSD~\cite{Song:2015sfa}								& 103.2$/$15		& 191.4$/$15 	& 294.7$/$30 \\
POWLANG-HTL~\cite{Beraudo:2014boa,Beraudo:2017gxw}	& 331.0$/$15		& 137.6$/$15  	& 468.6$/$30 \\
TAMU~\cite{He:2019vgs}								& 16.7$/$15		& 13.5$/$15  	& 30.2$/$30  \\
\bottomrule
\end{tabular}
\label{tab:chi2}	
\end{table}

\begin{figure}[!t]
\begin{center}
\includegraphics[width=1\textwidth]{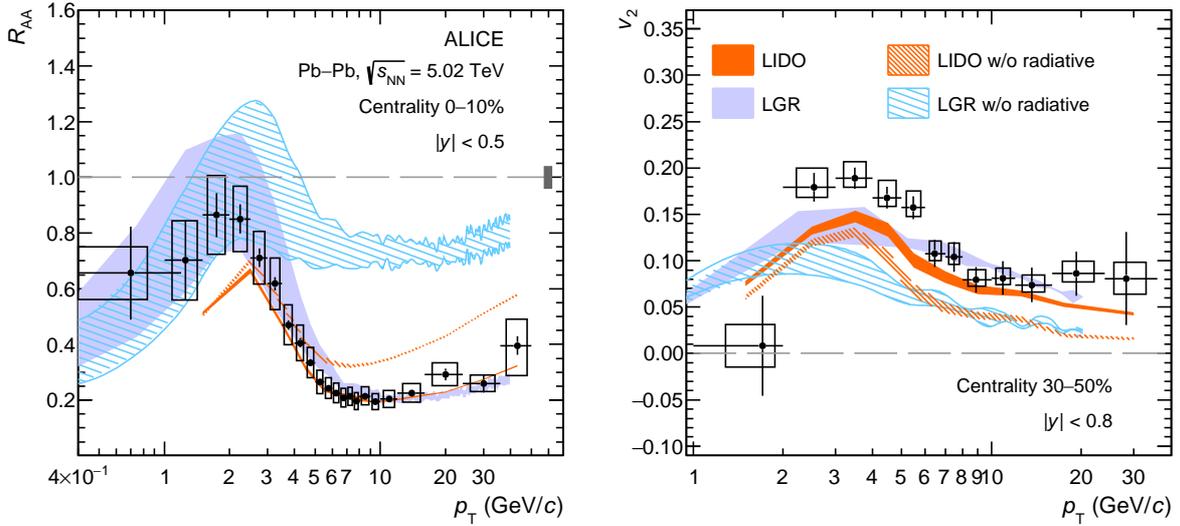}
\caption{Prompt D-meson $\RAA$ in the 0--10\% centrality class (left panel) and $v_{2}$ in the 30--50\% centrality class (right panel) compared with the LIDO~\cite{Ke:2018jem} and LGR~\cite{Li:2019lex} predictions obtained with and without including radiative processes in the  charm-quark interactions with the medium.} 
\label{fig:Dmeson_collrad}
\end{center}
\end{figure}

\begin{figure}[!t]
\begin{center}
\includegraphics[width=1\textwidth]{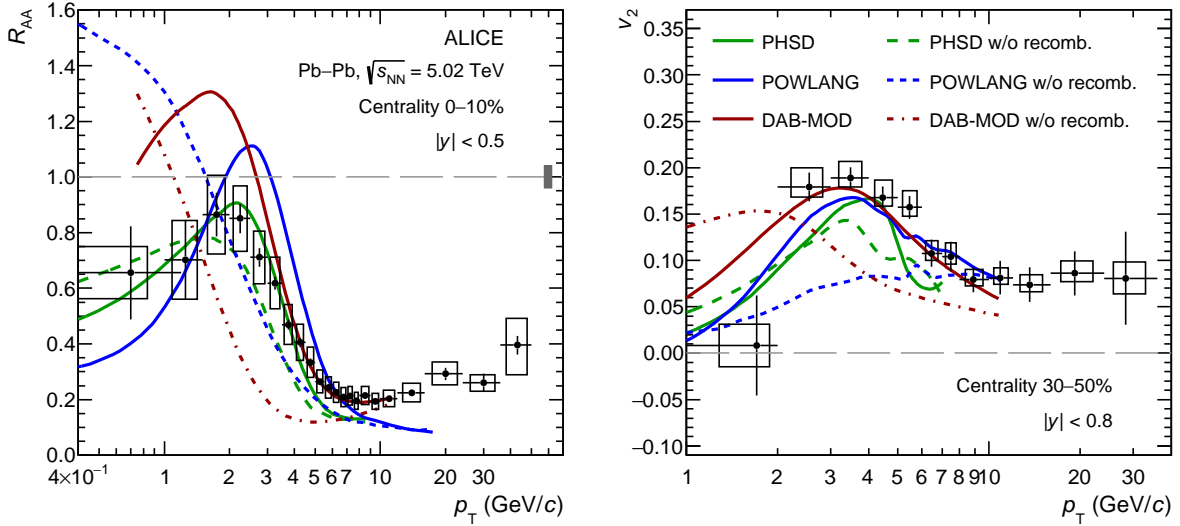}
\caption{Prompt D-meson $\RAA$ in the 0--10\% centrality class (left panel) and $v_{2}$ in the 30--50\% centrality class (right panel) compared with the 
PHSD~\cite{Song:2015sfa}, POWLANG~\cite{Beraudo:2014boa,Beraudo:2017gxw}, and DAB-MOD~\cite{Katz:2019fkc} predictions obtained with and without including hadronisation via recombination.} 
\label{fig:Dmeson_fragcoal}
\end{center}
\end{figure}

A deeper insight on the impact of the different implementations of the charm-quark interaction and hadronisation in the QGP can be obtained from Figs.~\ref{fig:Dmeson_collrad} and~\ref{fig:Dmeson_fragcoal}.
In Fig.~\ref{fig:Dmeson_collrad} the $\RAA$ and the $v_2$, measured in the 0--10\% and 30--50\% centrality classes, respectively, are compared with two different implementations of the LIDO~\cite{Ke:2018jem} and LGR~\cite{Li:2019lex} models, in order to assess the role of elastic and radiative processes in the charm-quark interactions with medium constituents. 
In particular, the first implementation is the standard one including both elastic and radiative processes, while the second prediction was obtained by switching off the radiative processes. At low $\pt$ (i.e. $\lesssim$ 3--4$~\gevc$), the collisional processes are expected to be the dominant interaction and this is confirmed by the similarity of the predictions for prompt D-meson $\RAA$ and $v_2$ with (full band) and without (striped band) radiative processes. This is further supported by the agreement, in the same $\pt$ range, between the experimental data and other theoretical models which implement only elastic processes, like TAMU~\cite{He:2019vgs}, PHSD~\cite{Song:2015sfa}, and POWLANG-HTL~\cite{Beraudo:2014boa,Beraudo:2017gxw}. Radiative processes, instead, become dominant at intermediate and high $\pt$. This can be observed in Fig.~\ref{fig:Dmeson_collrad} where the predictions without these processes overestimate (underestimate) the measured $\RAA$ ($v_2$) for $\pt \gtrsim$ 5--6$~\gevc$.

Similarly, Fig.~\ref{fig:Dmeson_fragcoal} shows the comparison of the experimental data for $\RAA$ and $v_2$ with two different versions of the 
PHSD~\cite{Song:2015sfa}, POWLANG-HTL~\cite{Beraudo:2014boa,Beraudo:2017gxw}, and DAB-MOD~\cite{Katz:2019fkc} models, in order to investigate the effects of the hadronisation mechanism, and in particular the role of recombination. This plays a key role in the predictions for $\RAA$ and $v_2$, since the relation between the momentum of the charm hadron and that of the parent charm quark is different for the fragmentation process, where the hadron inherits a fraction of the initial quark momentum, and the recombination, where the D-meson $\pt$ is larger than the one of the charm quark and the charm hadron inherits also the collective flow of the light quark.
In Fig.~\ref{fig:Dmeson_fragcoal}, the predictions are provided with (solid line) and without (dashed line) the implementation of the recombination process in the hadronisation mechanism. The calculations performed including only the fragmentation process underestimate both the $\RAA$ and $v_2$, while the inclusion of the recombination of charm quarks with light quarks pushes the predictions closer to the experimental data. This indicates that recombination with light quarks from the medium plays a relevant role in the hadronisation of charm quarks at the QGP phase boundary.

\section{Conclusions}
\label{sec:conclusions}
We have reported the measurement of the $\pt$-differential production yields of prompt $\Dzero$, $\Dplus$, and $\Dstar$ mesons and charge conjugates at midrapidity ($|y| <$ 0.5) in central and semicentral \PbPb collisions at a centre-of-mass energy per nucleon pair $\sqrtsNN = 5.02~\TeV$. The results were obtained with the data sample collected at the end of 2018 with the ALICE detector. The $\pt$-spectra were measured in finer $\pt$ intervals with respect to the previous measurements at the same centre-of-mass energy~\cite{Acharya:2018hre}, providing a more precise description of the $\pt$ distribution. 
The large data sample allowed for the first measurement of the $\Dzero$-meson yield down to $\pt =$ 0 in \PbPb collisions at the LHC. This enabled the determination of the $\pt$-integrated production yields of prompt $\Dzero$, which is obtained in a model independent way, of prompt $\Dplus$ and $\Dstar$, and the comparison with predictions from the statistical hadronisation model~\cite{Andronic:2019wva,Andronic:2021wva}.
The average $\RAA$ of the $\Dzero$, $\Dplus$, and $\Dstar$ mesons reaches a maximum suppression value of 5.5 (i.e. $\RAA\sim$ 0.18) and 2.7 (i.e. $\RAA\sim$ 0.4) in the 0--10\% and 30--50\% centrality classes, respectively, at \mbox{$\pt =$ 6--8$~\gevc$}. This suppression becomes less pronounced in peripheral collisions with a minimum value of 0.7--0.8, as observed in the 60--80\% centrality class in Ref.~\cite{Acharya:2018hre}, and it is due to final-state effects, since it is not observed in minimum bias \pPb collisions~\cite{Acharya:2019mno}. However, it was pointed out in Ref.~\cite{Loizides:2017sqq, ALICE:2018ekf} that event selection and geometry biases can cause an apparent suppression of $\RAA$, especially in peripheral collisions, even in the absence of nuclear effect.
The $\RAA$ of prompt D mesons at $\pt >$ 8$~\gevc$ is well described by models which include both collisional and radiative energy loss processes.
In addition, the $\pt$-integrated $\RAA$ of prompt $\Dzero$ mesons was obtained and compared with the $\RpPb$ measured in \pPb collisions at the same centre-of-mass energy~\cite{Acharya:2019mno}. The results lie on the upper edge of the calculations which consider nuclear modification of the PDFs.

In central collisions, the $\pt$-differential $\RAA$ of prompt D mesons is compatible within uncertainties with that of charged particles for $\pt \gtrsim$ 8 $\gevc$ and prompt $\Jpsi$ mesons for $\pt \gtrsim$ 10 $\gevc$, while an ordering in $\RAA$ is observed at lower $\pt$. The latter is due to the interplay of different effects, such as the diffusion of charm quarks in the medium possibly leading to their thermalisation in the QGP and the hadronisation via recombination with light quarks from the medium, which along with energy loss contribute to the modification of the $\pt$-distribution of the hadrons. The comparison of the $\RAA$ of prompt D mesons and non-prompt $\Jpsi$ for $\pt >$ 6$~\gevc$ together with model calculations supports the predictions of quark-mass dependent energy loss.
At low and intermediate $\pt$ ($\lesssim$ 6--8$~\gevc$), the $\RAA$ is well described by different transport model calculations for the two centrality classes. However, most of them fail in describing simultaneously both $\RAA$ and $v_2$ in central and semicentral \PbPb collisions. By considering the few models that are in fair agreement with both observables, the heavy-quark spatial diffusion coefficient was estimated to be in the range $1.5 < 2\pi D_\mathrm{s} T_\mathrm{c} < 4.5$ at the pseudocritical temperature $T_\mathrm{pc} = 155~\MeV$, which is a narrower interval as compared to estimations based on previous D-meson measurements at LHC energies~\cite{Acharya:2017qps,Acharya:2020pnh}.
Therefore, the simultaneous comparison of the data with the theoretical predictions for these observables allowed for more stringent constraints to the heavy-quark spatial diffusion coefficient and for significant progress in the understanding of the interaction processes and the hadronisation mechanisms of charm quarks in the high-density QCD medium. 


\newenvironment{acknowledgement}{\relax}{\relax}
\begin{acknowledgement}
\section*{Acknowledgements}

The ALICE Collaboration would like to thank all its engineers and technicians for their invaluable contributions to the construction of the experiment and the CERN accelerator teams for the outstanding performance of the LHC complex.
The ALICE Collaboration gratefully acknowledges the resources and support provided by all Grid centres and the Worldwide LHC Computing Grid (WLCG) collaboration.
The ALICE Collaboration acknowledges the following funding agencies for their support in building and running the ALICE detector:
A. I. Alikhanyan National Science Laboratory (Yerevan Physics Institute) Foundation (ANSL), State Committee of Science and World Federation of Scientists (WFS), Armenia;
Austrian Academy of Sciences, Austrian Science Fund (FWF): [M 2467-N36] and Nationalstiftung f\"{u}r Forschung, Technologie und Entwicklung, Austria;
Ministry of Communications and High Technologies, National Nuclear Research Center, Azerbaijan;
Conselho Nacional de Desenvolvimento Cient\'{\i}fico e Tecnol\'{o}gico (CNPq), Financiadora de Estudos e Projetos (Finep), Funda\c{c}\~{a}o de Amparo \`{a} Pesquisa do Estado de S\~{a}o Paulo (FAPESP) and Universidade Federal do Rio Grande do Sul (UFRGS), Brazil;
Ministry of Education of China (MOEC) , Ministry of Science \& Technology of China (MSTC) and National Natural Science Foundation of China (NSFC), China;
Ministry of Science and Education and Croatian Science Foundation, Croatia;
Centro de Aplicaciones Tecnol\'{o}gicas y Desarrollo Nuclear (CEADEN), Cubaenerg\'{\i}a, Cuba;
Ministry of Education, Youth and Sports of the Czech Republic, Czech Republic;
The Danish Council for Independent Research | Natural Sciences, the VILLUM FONDEN and Danish National Research Foundation (DNRF), Denmark;
Helsinki Institute of Physics (HIP), Finland;
Commissariat \`{a} l'Energie Atomique (CEA) and Institut National de Physique Nucl\'{e}aire et de Physique des Particules (IN2P3) and Centre National de la Recherche Scientifique (CNRS), France;
Bundesministerium f\"{u}r Bildung und Forschung (BMBF) and GSI Helmholtzzentrum f\"{u}r Schwerionenforschung GmbH, Germany;
General Secretariat for Research and Technology, Ministry of Education, Research and Religions, Greece;
National Research, Development and Innovation Office, Hungary;
Department of Atomic Energy Government of India (DAE), Department of Science and Technology, Government of India (DST), University Grants Commission, Government of India (UGC) and Council of Scientific and Industrial Research (CSIR), India;
Indonesian Institute of Science, Indonesia;
Istituto Nazionale di Fisica Nucleare (INFN), Italy;
Japanese Ministry of Education, Culture, Sports, Science and Technology (MEXT), Japan Society for the Promotion of Science (JSPS) KAKENHI and Japanese Ministry of Education, Culture, Sports, Science and Technology (MEXT)of Applied Science (IIST), Japan;
Consejo Nacional de Ciencia (CONACYT) y Tecnolog\'{i}a, through Fondo de Cooperaci\'{o}n Internacional en Ciencia y Tecnolog\'{i}a (FONCICYT) and Direcci\'{o}n General de Asuntos del Personal Academico (DGAPA), Mexico;
Nederlandse Organisatie voor Wetenschappelijk Onderzoek (NWO), Netherlands;
The Research Council of Norway, Norway;
Commission on Science and Technology for Sustainable Development in the South (COMSATS), Pakistan;
Pontificia Universidad Cat\'{o}lica del Per\'{u}, Peru;
Ministry of Education and Science, National Science Centre and WUT ID-UB, Poland;
Korea Institute of Science and Technology Information and National Research Foundation of Korea (NRF), Republic of Korea;
Ministry of Education and Scientific Research, Institute of Atomic Physics and Ministry of Research and Innovation and Institute of Atomic Physics, Romania;
Joint Institute for Nuclear Research (JINR), Ministry of Education and Science of the Russian Federation, National Research Centre Kurchatov Institute, Russian Science Foundation and Russian Foundation for Basic Research, Russia;
Ministry of Education, Science, Research and Sport of the Slovak Republic, Slovakia;
National Research Foundation of South Africa, South Africa;
Swedish Research Council (VR) and Knut \& Alice Wallenberg Foundation (KAW), Sweden;
European Organization for Nuclear Research, Switzerland;
Suranaree University of Technology (SUT), National Science and Technology Development Agency (NSDTA) and Office of the Higher Education Commission under NRU project of Thailand, Thailand;
Turkish Energy, Nuclear and Mineral Research Agency (TENMAK), Turkey;
National Academy of  Sciences of Ukraine, Ukraine;
Science and Technology Facilities Council (STFC), United Kingdom;
National Science Foundation of the United States of America (NSF) and United States Department of Energy, Office of Nuclear Physics (DOE NP), United States of America.    
\end{acknowledgement}

\bibliographystyle{utphys}   
\bibliography{dmeson_raa}

\newpage
\appendix
\section{The ALICE Collaboration}
\label{app:collab}
\small
\begin{flushleft} 

\bigskip 

S.~Acharya$^{\rm 142}$, 
D.~Adamov\'{a}$^{\rm 97}$, 
A.~Adler$^{\rm 75}$, 
J.~Adolfsson$^{\rm 82}$, 
G.~Aglieri Rinella$^{\rm 34}$, 
M.~Agnello$^{\rm 30}$, 
N.~Agrawal$^{\rm 54}$, 
Z.~Ahammed$^{\rm 142}$, 
S.~Ahmad$^{\rm 16}$, 
S.U.~Ahn$^{\rm 77}$
% ...

\bigskip

\bigskip 

\textbf{\Large Affiliation Notes}

\bigskip 

% ...
$^{\rm VII}$ Also at: University of Kansas, Lawrence, Kansas, United States\\

\bigskip

\bigskip 

\textbf{\Large Collaboration Institutes}

\bigskip 

% ...
$^{16}$ Department of Physics, Aligarh Muslim University, Aligarh, India\\
% ...
$^{30}$ Dipartimento DISAT del Politecnico and Sezione INFN, Turin, Italy\\
% ...
$^{34}$ European Organization for Nuclear Research (CERN), Geneva, Switzerland\\
% ...
$^{54}$ INFN, Sezione di Bologna, Bologna, Italy\\
% ...
$^{75}$ Johann-Wolfgang-Goethe Universit\"{a}t Frankfurt Institut f\"{u}r Informatik, Fachbereich Informatik und Mathematik, Frankfurt, Germany\\
% ...
$^{77}$ Korea Institute of Science and Technology Information, Daejeon, Republic of Korea\\
% ...
$^{82}$ Lund University Department of Physics, Division of Particle Physics, Lund, Sweden\\
% ...
$^{97}$ Nuclear Physics Institute of the Czech Academy of Sciences, \v{R}e\v{z} u Prahy, Czech Republic\\
% ...
$^{142}$ Variable Energy Cyclotron Centre, Homi Bhabha National Institute, Kolkata, India\\
% ...

\bigskip 

\end{flushleft} 
\end{document}